# STFC Astronomy Advisory Panel Roadmap 2022

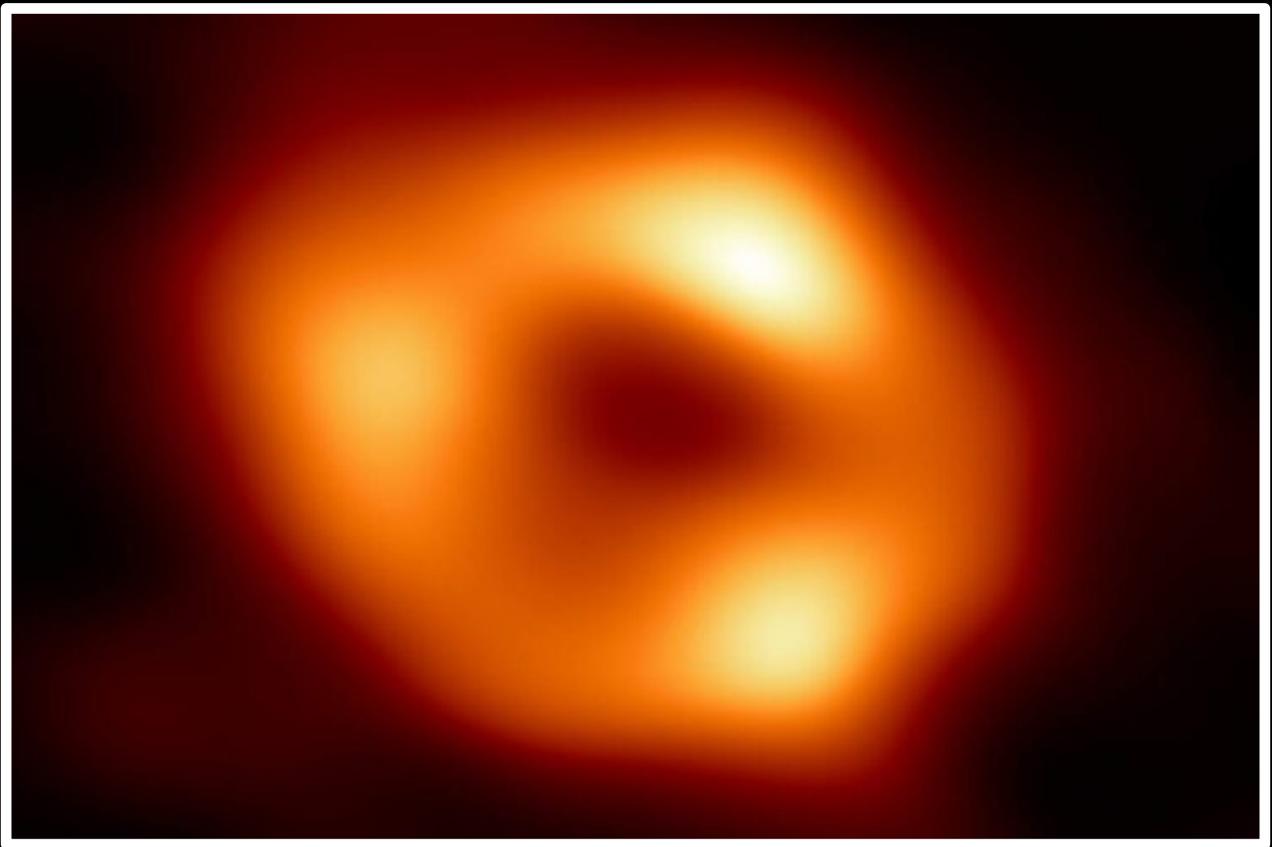

Stephen Serjeant (Open University, chair), James Bolton (University of Nottingham), Poshak Gandhi (University of Southampton), Ben Stappers (University of Manchester), Paolo Mazzali (Liverpool John Moores University), Aprajita Verma (University of Oxford), Noelia E. D. Noël (University of Surrey)

# STFC Astronomy Advisory Panel Roadmap 2022

**Astronomy Advisory Panel:** Stephen Serjeant (Open University, chair), James Bolton (University of Nottingham), Poshak Gandhi (University of Southampton), Ben Stappers (University of Manchester), Paolo Mazzali (Liverpool John Moores University), Aprajita Verma (University of Oxford), Noelia E. D. Noël (University of Surrey)

*The cover image shows the first image of Sgr A\*, the supermassive black hole at the centre of our Galaxy[1,2,3,4,5,6,7]. UK astronomers are part of the Event Horizon Telescope team that made this image. The observatories involved include the James Clerk Maxwell Telescope, with UK access funded by PPRP and UK university contributions, and ALMA, where UK access is through ESO. UK ALMA users in particular are also supported by the STFC-funded UK ALMA Regional Centre. Image credit: Event Horizon Telescope collaboration.*

---

[1] See https://www.ukri.org/news/first-image-of-black-hole-in-the-centre-of-our-galaxy-unveiled/

[2] The Event Horizon Telescope Collaboration. 5/12/2022. "First Sagittarius A\* Event Horizon Telescope Results. I. The Shadow of the Supermassive Black Hole in the Center of the Milky Way." The Astrophysical Journal Letters, 930, L12. https://doi.org/10.3847/2041-8213/ac6674

[3] The Event Horizon Telescope Collaboration. 5/12/2022. "First Sagittarius A\* Event Horizon Telescope Results. II. EHT and Multiwavelength Observations, Data Processing, and Calibration ." The Astrophysical Journal Letters, 930, L13. https://10.0.15.7/2041-8213/ac6675

[4] The Event Horizon Telescope Collaboration. 5/12/2022. "First Sagittarius A\* Event Horizon Telescope Results. III. Imaging of the Galactic Center Supermassive Black Hole ." The Astrophysical Journal Letters, 930, L14. https://doi.org/10.3847/2041-8213/ac6429

[5] The Event Horizon Telescope Collaboration. 5/12/2022. "First Sagittarius A\* Event Horizon Telescope Results. IV. Variability, Morphology, and Black Hole Mass." The Astrophysical Journal Letters, 930, L15. https://doi.org/10.3847/2041-8213/ac6736

[6] The Event Horizon Telescope Collaboration. 5/12/2022. "First Sagittarius A\* Event Horizon Telescope Results. V. Testing Astrophysical Models of the Galactic Center Black Hole." The Astrophysical Journal Letters, 930, L16. https://doi.org/10.3847/2041-8213/ac6672

[7] The Event Horizon Telescope Collaboration. 5/12/2022. "First Sagittarius A\* Event Horizon Telescope Results. VI. Testing the Black Hole Metric ." The Astrophysical Journal Letters, 930, L17. https://doi.org/10.3847/2041-8213/ac6756

# STFC Astronomy Advisory Panel Roadmap 2022

Astronomy Advisory Panel: Stephen Serjeant (Open University, chair), James Bolton (University of Nottingham), Poshak Gandhi (University of Southampton), Ben Stappers (University of Manchester), Paolo Mazzali (Liverpool John Moores University), Aprajita Verma (University of Oxford), Noelia E. D. Noël (University of Surrey)

# Contents

# 1. Executive Summary

These science and technology roadmaps provide a brief overview of UK Astrophysics research, areas of strength, opportunities for growth and areas for strategic investment. Observational and theoretical astronomy are fast-paced fields in which progress depends on a broad range of long-term, large-scale facility investments, from international observatories to high-performance computing. UK astronomy performs exceptionally well internationally on a wide variety of metrics (for example: for refereed astronomy papers published in 2000-2020, those with UK co-authors received 2.7M citations[8], compared to e.g. 7.1M co-authored from the USA[9] from a community approximately 4.2 times

---

[8] Source: NASA Astrophysics Data System
[9] Source: NASA Astrophysics Data System

larger[10]), at least some of which may be attributable to strategic positioning within a wide range of long-term international projects.

The Astronomy Advisory Panel have therefore consulted with the UK community to compile roadmaps that capture at least some of the breadth and depth of the current community aspirations. However, it is equally important to emphasise what these roadmaps are not. Firstly, in a significant change to the advisory panel remit since the last major roadmap refresh in 2012, the tensioning between current spending priorities is now explicitly conducted through the advice of Balance of Programmes reviews (BoP), the last of which[11] was in 2020. The roadmaps are not intended as a substitute for this process (or any potential astronomy-specific equivalent), but rather to highlight current aspirations for hypothetical future spends. Secondly, the fast-paced nature of the field implies that the aspirations are subject to change. For example, some of our community consultation responses were explicit that it is impossible to identify unambiguously the technologies needed to address STFC's key science challenges in the coming ten years, and that therefore the only defensible strategic choice that maximises the probable science return is to favour the breadth of the technology programme. Finally, these roadmaps must not be used as the sole justification for the allocation of new funding resources, without discussion with advisory panels about the consequences for other parts of the programme and consultation on any subsequent developments in astronomy.

The top priority of the UK astronomy community remains exploitation grants, despite a very welcome recent uplift of £2M year on year (leading to a £6M increase by 2024/25). This priority has been consistent over the past ten years. Flat-cash STFC funding, combined with the increasing costs of facility funding required to maintain a world-class astronomy programme, has resulted in a radical erosion of PDRA funding. At current levels, a typical research-active academic might secure postdoc funding three times in an entire career. Not only is this a bottleneck for science exploitation, it drives an unplanned and potentially inequitable attrition of early-career talent from UK astrophysics. There is evidence presented in the roadmaps that the community view of the balance of funding is evolving in response to this chronic underfunding of exploitation.

Exploitation funding aside, UK astronomy has benefitted from long-term strategic facility and technology funding that has driven a very wide range of high-profile science results. Unlike for example particle physics, astronomy has a wide and diverse range of science goals from the microscopic processes in astrochemistry, to exoplanetary science, to star formation and evolution, to the evolution of galaxy populations and the large scale structure of the Universe, to the properties of dark matter and dark energy, to the fundamental cosmological parameters and theories of gravity, and more, all driven by UK access to world-leading facilities, most of which address a very broad range of these science goals. The diversity in this portfolio of facilities is inevitable and essential, and is driven in part by the wide range of physical processes that dominate in the various parts of the electromagnetic spectrum. Currently, the UK astronomy community aspirations continue to follow a similar pattern of strongly supporting these long-term investments, and have led to **a very clear set of recommendations for sustaining a vigorous timeline of UK astronomy capabilities.**

Among the top community priorities is the **Square Kilometer Array (SKA)**, which is supported by a very broad section of the community and which promises transformative progress across a very wide range of STFC priority areas. A high community priority is the UK role in the **Low Frequency Array (LOFAR)**, which as a pathfinder to the SKA is already producing world-leading survey results, and in which the UK is very well positioned for future success. The UK involvement in the **European Southern Observatory (ESO)** is critical to UK astronomy, and is another of our top community priorities. ESO's **Very Large Telescopes** (VLT) provides critical capability across almost the entirety of the UK astronomy programme. ESO's **Extremely Large Telescope (ELT)** is a top priority for the long term UK astronomy capability and for nearer-term technology development, and will clearly revolutionise astronomy throughout the UK programme. Through ESO, the UK also has access to the ~£1.3Bn **Atacama Large Millimetre Array (ALMA)** in which the UK submm/mm-wavelength community have capitalised on their strong heritage in securing competitive allocations, supported by the UK ALMA Regional Centre and with many programmes fed by the UK role in the **James Clerk Maxwell Telescope (JCMT)** (funded via PPRP and university contributions). The UK also plays leading roles in building facilities and/or associated instruments (e.g. ELT, VLT, SKA, ALMA, etc.), often strategically placing the UK to trade unique technical capabilities, often mission-critical, for a seat at the table setting the scientific agenda of a mission.

---

[10] Source: International Astronomical Union, https://www.iau.org/public/themes/member_statistics/
[11] See https://www.ukri.org/wp-content/uploads/2022/03/STFC-210322-BalanceOfProgrammeExercise2020.pdf

UKSA is responsible for much of the space mission programme, while the science strategy, early technology, data challenges and exploitation are all within the remit of STFC; nevertheless, the roadmaps would be incomplete if they failed to note that the **ESA Science mandatory programme** covers projects supporting almost the entirety of the UK astronomy science and technology communities, with missions comparable (or greater) in cost, impact and community support to our top-rated ground-based projects.

There is a very strong track record in astronomy of major discoveries following the opening of new astronomical parameter spaces. Time-domain astronomy is an excellent topical example, not just with follow-ups of gravitational wave events (see PAAP roadmaps), fast radio bursts, gamma-ray bursts, supernovae and other transients, but also in the explosion of interest and UK world-leadership in the field of exoplanets. The **Vera Rubin Observatory Legacy Survey of Space and Time (Rubin LSST)** will image the whole available sky twice a week for ten years at optical wavelengths, and is one of the community's clear Very High priorities.

A further Very High priority driven by the UK's world-leading cosmic microwave background (CMB) community is the UK role in the **Simons Observatory**, which has been funded outside the STFC core programme through a UKRI infrastructure call. This is the top priority of the UK CMB community and builds on a very strong heritage of international leadership in CMB cosmology. Also in the Very High priority science category, and Very High in technology priority (our highest categorisation for technology) is UK capability in **High Performance Computing (HPC)**, which underpins a wide range of theoretical astrophysics from planet formation to cosmology.

There is also support for many smaller facility projects. Here, 'small' should not be equated with 'not world-leading'; there is no place in these roadmaps for any facility that is not world leading. While these projects may not enjoy the scientific breadth and almost-universal community involvement of our top priorities, they nevertheless offer the prospect of transformative progress in particular areas. In many cases it is not enough for there to be only access to the largest international facilities; national access to world-leading smaller specialist facilities can provide feeder programmes to the larger facilities and provide the UK with tactical and strategic advantages (e.g. JCMT feeding ALMA and e-MERLIN feeding SKA). See Figure 1.

In summary, astronomy in all its forms is a UK strength. The UK strength is across the breadth of the field, from planetary science to cosmology. There is a clear consensus on the key science questions to be addressed, and these require access to a wide range of facilities and capabilities, many of which address multiple key science questions. International partnerships (e.g. ESA, ESO, SKA and bilateral programmes) are extremely important in order for the UK to play leading roles in international science teams, including across Europe. A strong relationship with UKSA and ESA, and engagement with their plans, is also essential. The UK astronomy programme is naturally split across several STFC advisory panels (e.g. AAP, SSAP, PAAP), but there are many synergies, in terms of the need for access to exploitation funds, the support for theoretical studies, training & skills, computing resources, instrumentation and technology development, etc. The advisory panels are therefore not in competition but rather are synergistic, and where there are overlapping interests there is concordance over the science and technology aspirations.

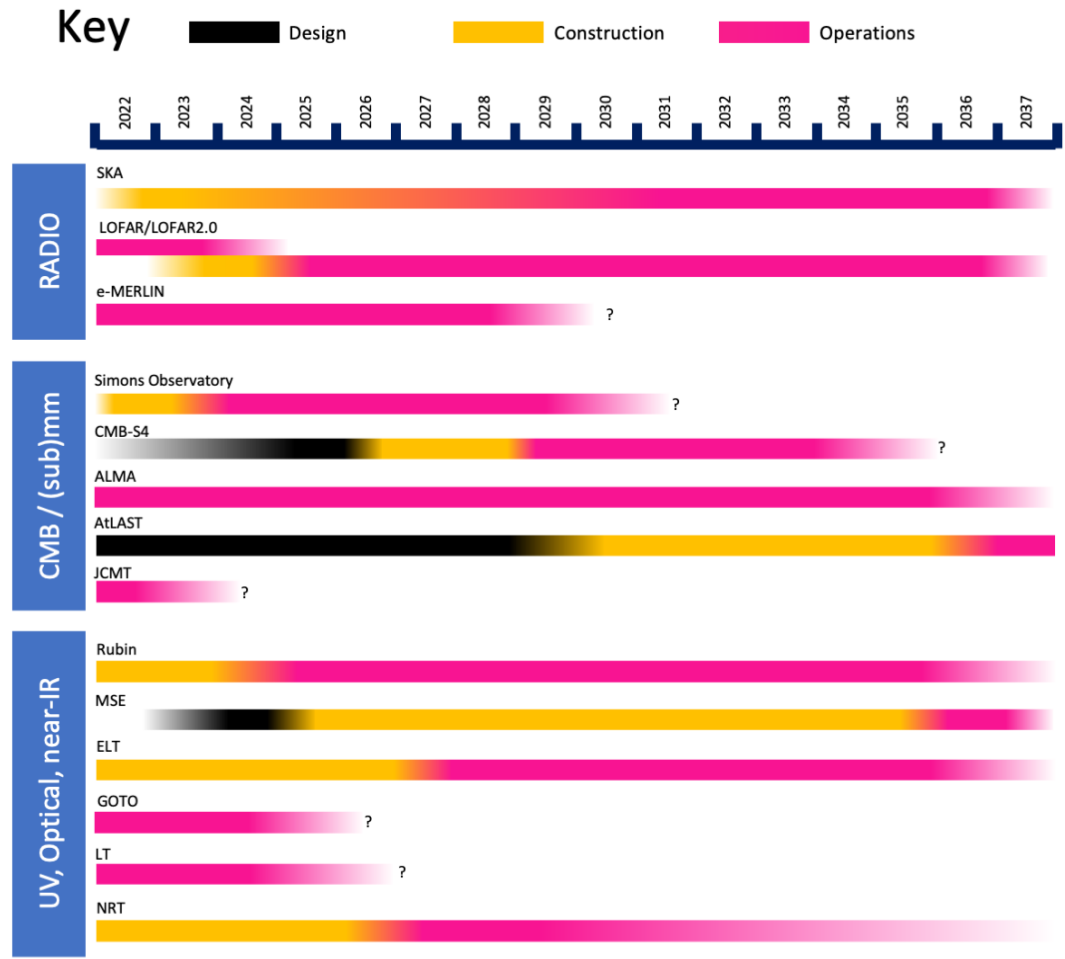

**Figure 1:** Schematic Gantt chart of a selection of significant facilities[12,13,14,15,16,17,18,19,20,21] featured in the AAP Roadmap. Timescales are indicative only and dependent on a wide variety of factors. Project end dates do not indicate AAP decisions but in many cases reflect current review dates at the time of writing. Only the facilities themselves are shown, and not individual instruments (e.g. WEAVE).

---

[12] SKA timeline adapted from https://www.skao.int/en/explore/construction-journey and https://www.skao.int/en/science-users/159/scientific-timeline

[13] LOFAR 2.0 timescales from Philip Best, priv. comm., November 2022: Nov 7th 2022: Formal sign-off by International LOFAR Telescope Board on procurement contracts; March 2023: delivery of equipment for three stations; March 2024: system commissioning/testing with 3-station array; Sept 2024 to mid-2025: full system roll-out.

[14] e-MERLIN timescales of existing funding from Simon Garrington, priv. comm., November 2022; there is no horizon for the ERIC as contributions are expected to be renewed by the current partners and new partners are being sought.

[15] Rubin timeline adapted from https://www.lsst.org/about/project-status

[16] AtLAST timescales based on https://arxiv.org/abs/2011.07974 with a purely illustrative ~6-year construction (cf. approximately 4 years for JCMT, 8 for ALMA to achieve early science, approximately 9 for LMT)

[17] JCMT timescale based on currently awarded STFC funding, without precluding further funding in future https://gtr.ukri.org/projects?ref=ST%2FV000268%2F1

[18] For Simons Observatory and CMB-S4, see PAAP roadmap

[19] GOTO, LT, NRT: adapted from indicative timescales and current funding commitments from C. Vincent, priv. comm., November 2022, without precluding further future funding

[20] MSE: timescales adapted from White Paper submitted to AAP

[21] ELT timescales adapted from https://elt.eso.org/about/timeline/

# 2. International context and consultation process

## 2.1 Background of previous AAP reviews

AAP conducted a call for white papers and consultation exercise with the UK astronomy community via an online survey in November 2021. This was advertised via astrocommunity and other routes. We greatly appreciate the help of our STFC colleagues in assembling the survey, providing us with the results promptly in a digestible format and in providing other contextual information. The results of the consultation are described in detail elsewhere[22], and summarised in Section 3 below.

The AAP science roadmap last had a major update[23] in 2012, and the technology roadmap is even further outdated. In the meantime, there have been minor updates following AAP's 2016 roadmap consultation[24], and its 2018 consultation to feed into the second STFC Balance of Programmes review[25]. The non-confidential parts of AAP's 2018 consultation report were published[26] in Astronomy and Geophysics.

AAP nevertheless opted to delay its most recent consultation and report for several reasons. Firstly, the global health crisis meant that the capacities of both the panel and the wider community were limited. Secondly, the outcomes of the European ASTRONET review could affect the policy decisions for AAP. Finally, the US Decadal Review outcomes would also be likely to affect the AAP roadmap. Both the ASTRONET and US Decadal reviews were also delayed partly due to the global health crisis. AAP do not seek to duplicate these roadmaps, but rather present the UK-specific community voice.

## 2.2 US Decadal Review 2020-1: priorities and synergies with UK interests

The US Decadal Review is a comprehensive survey of US community aspirations in both ground-based and space-based astronomy for not just the coming decade but also outlining a roadmap for future decades. In ground-based astronomy, the review prioritises participation in extremely large telescopes (30 metre class), the next generation Stage 4 cosmic microwave background (CMB) experiments, the next generation Very Large Array radio telescope ngVLA. Unlike the UK, the US is not a participant in the Square Kilometer Array (SKA); instead the focus is on US-specific complementary higher frequency Northern hemisphere capabilities in ngVLA. The US prioritisation of extremely large telescopes is very similar to the UK community's prioritisation of science opportunities and technical developments in the European Southern Observatory's Extremely Large Telescope, ELT. The coming decades of ground-based optical and near-infrared astronomy will clearly be dominated internationally by facilities of this class. The CMB aspirations are very well aligned with those in the UK, and UK participation in the US-led Simons Observatory is now funded (see below). The review also prioritised the Vera Rubin Observatory, which also features as a UK very high priority (see below). US priorities in astroparticle physics include upgrading the neutrino detector IceCube and gravitational wave technology development, though in UK terms these are overlapping with PAAP interests. The review calls for an expansion in grants (referred to as the exploitation line in STFC terminology), supporting data archives and curation, bolstering theory underpinnings, advancing laboratory measurements (e.g. laboratory astrochemistry), and expanding basic technology development. Most of these items feature in the high or very high UK community priorities discussed below. The US prioritisation of data archives and curation echoes a European prioritisation of open science under the "FAIR" principles: Findable, Accessible, Interoperable and Reusable. However, in the UK our consultation highlighted not only a need for continued support for this area, but rather a conspicuous lack of support for software, archiving, instrumentation and technical careers. The US is prioritising an expansion in grants, while the UK has for many years been failing to make optimal use of its astronomy facility investments due to a chronic underfunding of the exploitation line.

---

[22] See https://www.ukri.org/publications/astronomy-advisory-panel-summary-of-2021-community-consultation/
[23] No longer available on the UKRI websites, but archived at https://www.yumpu.com/en/document/view/4206082/aap-prreport-submitted-nov22
[24] No longer available on the UKRI websites, but archived at https://web.archive.org/web/20220305101010/https://stfc.ukri.org/files/aap-balance-report/
[25] See https://www.ukri.org/about-us/stfc/planning-strategy-reviews/balance-of-programmes/
[26] See Serjeant, S., et al., 2019, Astronomy & Geophysics, Volume 60, Issue 2, Pages 2.13–2.17, https://academic.oup.com/astrogeo/article/60/2/2.13/5380734

In terms of overarching AAP-area science themes, the US Decadal Review recognises several key areas. "Pathways to Habitable Worlds" supports the work of the US exoplanet community, which has a sizable UK counterpart, and in terms of ground-based astronomy requires support from ELTs, resolution spectroscopy, high-performance adaptive optics, high-contrast imaging, as well as laboratory and theoretical studies, all of which have been recognised as UK community priorities below. The "New Messengers and New Physics" and "New Windows on the Dynamic Universe" themes cover not just CMB cosmology but also the rapidly expanding discovery space of time-domain astrophysics, including gravitational wave counterparts and other optical and radio transients, all of which have sizeable UK community support. Ground-based technical capabilities required to support these themes are very similar to those highlighted by the UK communities and include ELTs, next generation CMB experiments and gravitational wave technology developments. The "Cosmic Ecosystems" theme covers a wide range of galaxy evolution studies, requiring ground-based support from the Vera Rubin Observatory as well as a wide range of space-based facilities including the forthcoming ESA-led Euclid mission, while the "Unveiling the Drivers of Galaxy Growth" theme pushes also to higher redshifts with the help of ELTs, new radio facilities, and space missions. Both themes also recommend support for theoretical underpinnings. The UK has a vibrant extragalactic community and prioritises a similar mix of ground-based technical capabilities, though the SKA affords a slightly different and complementary set of opportunities. The consistent call for theoretical underpinning is reflected also in the UK community prioritisation of continuously updating the national high performance computing capability, discussed below.

The US Decadal Review also set out aspirations for space-based astronomy, the headlines of which are a large infrared/optical/ultraviolet space telescope, X-ray and far-infrared missions. Together, these missions cover the breadth of the review's science themes. In the UK, space mission involvements are the purview of UKSA and are formally outside the remit of STFC, except in some limited situations discussed briefly below. Nevertheless, it is also the case in the UK that space missions are central to a wide range of science themes, including the US priorities of JWST, Roman, ESA Euclid, as well as other current and future ESA M-class and L-class missions. The limitations of atmospheric stability, transmission and/or emissivity inevitably require space missions for many science goals, including exoplanet discovery, high energy astrophysics, time-domain astrophysics and physical processes obscured by dust. Although space projects are outside the AAP remit, and although AAP are not being asked for as fine-grained detail on community priorities for space missions as we are in ground-based astronomy, we felt it would be inappropriate if we did not capture the community view by ranking the UK involvement in the ESA mandatory science programme as being collectively equal to our highest ranked ground-based priorities, below. There is also a great deal of UK interest in space bilateral opportunities, including with the US missions highlighted in the Decadal Review, but this is largely beyond the remit of the AAP roadmaps.

The review also made a wide range of recommendations on societal impacts, career development, equity, diversity and inclusion. Many of these are also appropriate for UK contexts and are discussed in the context of our community consultation below.

## 2.3 European ASTRONET review 2021-2022: priorities and synergies with UK interests

### 2.3.1 Context and background

ASTRONET is a European consortium of funding agencies, research organisations and associated bodies that was formed in 2005 with initial funding support from the European Commission. It is now self-sustaining (i.e. without EC financial contributions). One of its principal objectives is the creation of a community-driven strategic plan and infrastructure roadmap, covering all of European astronomy. The remit covers not just the ground-based astronomy within the UK STFC AAP remit, but also solar system science and space-based facilities across all of European astronomy.

ASTRONET last published its science vision and infrastructure roadmap[27] in 2015, after which its focus has been on the implementation through a series of reports on the European coordination of research infrastructures; the integration of the European communities in mainstream astronomy; the coordination of national funding agencies; and education, training, and public outreach. Most recently, over the past two years ASTRONET has been working on a new Science Vision and Infrastructure Roadmap. Again the aspiration is for this to include all aspects of European astronomy to aid national and multi-national long-term astronomy planning. An extensive community consultation[28] on the drafts concluded in August 2022, and at the time of writing (November 2022) five of the eight science themes exist in a final form, while the remaining three are still draft reports. Below we summarise the recommendations in each theme and the UK relevance. Note the wide range of facilities and instruments that are named as supporting several science challenges. Many UK colleagues participated in the ASTRONET consultation; there are no obvious areas in which UK aspirations are in conflict with those of this wider European strategy, but there are areas of particular UK interest.

### 2.3.2 Origin and Evolution of the Universe

This science theme outlines the current standard cosmological models and the community aspirations for progress in the coming two decades. The report highlights a growing overlap or synergy between fundamental physics and cosmology, through the physics of the early Universe and tracers of the matter distribution around the epoch of recombination, as well as the large-scale distribution of galaxies at lower redshifts, both of which are areas that have broad UK active communities with world-leading track records. The UK had a representative in the working group that drafted this chapter. Key science questions in those areas include: "Are there deviations from the standard model of particle physics? Are there deviations from the standard cosmological model?" The emerging tension in measurements of the Hubble constant hints either at unknown systematics or new physics, while the accelerating expansion of the Universe inevitably signals new physics; the report poses further key questions: "Are there deviations from general relativity, and on what scales? What is the origin of the accelerated expansion?" A further inevitable signal of physics beyond the Standard Model is the evidence for dark matter, all of which is currently astronomical, and a further key question in this section is "What is the nature of dark matter?" The temperature and polarisation anisotropy of the CMB probes physics between the LHC energy range of $10^4$ GeV to Grand Unified Theory scales of $10^{16}$ GeV, and a further key science question highlighted in the report is "Can we identify specific observational signatures of inflation?" Finally, this section of the report covers the PAAP-remit area of gravitational waves, posing the key questions "What can gravitational waves observations reveal about dark energy, dark matter and modifications of gravity on cosmological scales?"

In each of these key question areas, critical ground-based and space-based facilities are identified. In terms of ground-based astronomy, the report highlights the importance of the SKA, Rubin, Simons

---

[27] See https://www.ASTRONET-eu.org/archives
[28] See https://www.ASTRONET-eu.org/forums/roadmap-community-consultation

Observatory (and forthcoming CMB Stage 4), all of which are very high UK community priorities. European CMB ground-based (e.g. QUBIC, QUIJOTE, LSPE-strip) and balloon-borne experiments (e.g. LSPE-swipe, BISOU) highlighted as pathfinders for larger space and ground-based facilities. Ground-based spectroscopic projects also feature prominently, such as the US-led DESI (with UK member institutions), the Japanese Subaru-PFS, the Mauna Kea Spectroscopic Explorer (MSE, recommended by AAP for UKRI infrastructure funding, see below), and the US-led MegaMapper. For probing the epoch of reionization via 21cm cosmology, the report highlights LOFAR (with UK involvement), MWA and the GMRT, and forthcoming facilities LOFAR2.0, MWA2, uGMRT, with SKA already under construction and HERA operational. We note that UK involvement in HERA has been considered for PPRP funding but so far has not been successful. Line intensity mapping also features as a cosmological probe, tracing the large-scale structure of atomic, molecular and/or ionic gas in the Universe, and requiring ground-based facility support from SKA, APEX/CONCERTO (with ESO involvement), CCAT (with UK involvement), and BINGO (with UK involvement). The report recommends investing in "small/medium-size facilities and new instruments on existing platforms (e.g., WEAVE, 4MOST)" as pathfinders for more ambitious wide-field facilities (e.g. MSE). Finally, the report chapter recommends investments in data processing, analysis and data science to accompany new instrumentation, in concordance with similar, broader recommendations from the US Decadal Review.

*2.3.3 Formation and Evolution of Galaxies*
At the time of writing, this chapter currently only exists in draft form. There is a very broad and active community in both the UK and more widely across Europe in this science area. Accordingly, the report chapter covers an extremely wide range of science questions and concomitant facilities. At very early cosmic epochs, the chapter outlines the aspirations for understanding the physics during the reionization epoch, including the sources of reionization, the topology of ionized gas, the properties of the first stars and galaxies, and the origins of gas in galaxies that fuel star formation. At lower redshifts, the report highlights aspirations for understanding critical processes driving galaxy evolution and the growth of stellar mass, including positive and negative feedback processes from starbursts and active nuclei, the dark matter halo processes (besides mass) that drive galaxy properties, the state of gaseous halos surrounding galaxies, the processes that drive star formation out of dense molecular gas, the processes that drive the star formation histories and morphological evolution of galaxies, the role of environments and the accretion physics around supermassive black holes. The report extends to science questions covering the Milky Way, including its accretion history (and what information can be recovered about it), the assembly of its halo, the abundance patterns of its first stars, and our very local environment.

The facility requirements and aspirations that accompany this vision are similarly broad. Many current and future space missions are prioritised, though in UK contexts these are broadly in the remit of UKSA rather than STFC. In terms of ground-based spectroscopy, the report chapter highlights multiplexed spectroscopy (4MOST, ESO VLT/MOONS, WEAVE, Subaru PFS, CFHT/Mauna Kea Spectroscopic Explorer, ESO spectroscopic facility), high-resolution spectroscopy (CUBES and ELT/ANDES and MOSAIC), integral field spectroscopy (e.g. VLT/MUSE and BlueMUSE), and 21cm spectroscopy and line intensity mapping (SKA, HERA). Many of these have UK key roles, including ESO instrumentation, VLT/MOONS, and WEAVE; the Mauna Kea Spectroscopic Explorer was among the AAP priorities for UKRI infrastructure funding (see below). Ground-based imaging capabilities are also prioritised with Rubin and ultimately ELT. The report supports the development of interferometers such as LOFAR and including endorsing the aspirations in the ALMA 2030 roadmap[29], as well as developments of the German/Spanish/French PdBI/NOEMA array and supporting new facilities such as the SKA, the US ngVLA (see above), and a future large single-dish submm/mm-wave telescope AtLAST (which featured in an AAP recommendation for future UKRI infrastructure funding, see below). High angular resolution studies are prioritised on both the ESO VLTs and the future ELT (e.g. VLT/MAVIS, ELT/MICADO, ELT/HARMONI). High angular resolution interferometry with ALMA (especially the long baseline extension), VLBI, VLTI, ngVLA, SKA and its precursors such as MeerKAT are all prioritised, as is Rubin LSST. The Cherenkov Telescope Array (CTA) is also an ASTRONET priority (in which there is UK technical interest in instrumentation and UK science interests mainly in the prospects for multi-messenger / multi-wavelength astronomy). Also highlighted are cosmological and extragalactic simulations, including Illustris, the EAGLE simulations, the Hestia project, and many others.

---
[29] See https://www.almaobservatory.org/en/publications/the-alma-development-roadmap/

*2.3.4 Formation and Evolution of Stars*
This is a wide-ranging science theme in which the UK also has large, vibrant research communities; the UK chaired the working group that drafted this chapter and also had representation among its membership. The science theme covers the birth, lifecycle and death of stars, the physics of stellar remnants and the cosmological applications of transient events. Key questions in the area of stellar birth include: "How do molecular clumps fragment? What effect does the local magnetic field have on collapse? How do stars accrete material? What sets the upper mass limit of a star? Is the initial mass function (IMF) universal? What is the relationship of the IMF to the core mass function? What sets the initial multiplicity of stars, and how does this impact the IMF? How do young stars and their disks set the properties of nascent planetary systems and their early evolution? What chemical pathways are important in the production of complex organics?" The evolution of stars also poses a number of key science questions, including: "What are the basic constituents of [stellar] matter? What happens to matter in extreme conditions? What external factors influence the conditions required for life on planets?" Finally, ASTRONET highlights a number of key science questions regarding the end-states of stellar evolution, including: "What are the progenitors of type Ia supernovae? What is the Galactic low-frequency GW background? What is the ultimate fate of planetary systems orbiting stellar remnants?" Neutron stars pose a wide range of key science questions relating to their equations of state, the origin of their magnetic fields and radio emission, their formation routes and interaction with their environments. Stellar-mass black holes are important tests of General Relativity and there are open science questions on their formation routes.

A wide range of ground-based instruments and facilities are prioritised to support these science goals, including ELT, Rubin LSST, SKA, WEAVE, 4MOST, VLT/MOONS, VLT/CUBES, VLTI, as well as smaller facilities specialising in time-domain science such as GOTO (prioritised by AAP in the previous "priority projects" process and with UK involvement now funded by PPRP) and BlackGEM (again with UK involvement). Multi-messenger astronomy features prominently in this theme. Existing facilities with unique capabilities are also prioritised, including ALMA, VLT/VLTI, the Spanish 10.4 metre telescope GTC, WHT, and the Italian 3.58-metre TNG telescope. Looking ahead, the report prioritises new ground-based capabilities with 8-10 metre class facilities such as VLT GRAVITY+ and VLT BlueMUSE, new highly multiplexed spectroscopic survey capabilities such as wider European participation in the Mauna Kea Spectroscopic Explorer or similar (which featured in the AAP prioritisation for UKRI infrastructure funding, see below), as well as highlighting the need for a large single-dish sub-mm facility such as AtLAST (again featuring in AAP's UKRI infrastructure funding shortlist).

*2.3.5 Formation and Evolution of Planetary Systems*
Once again this is an area with a vibrant UK research community, and the UK had a representative on the working group that drafted this chapter. This field is experiencing an extraordinary phase of discovery since the first exoplanets were discovered only three decades ago. The overarching science questions highlighted by ASTRONET that are driving the current and future research in this area are: "What drives the enormous diversity in exoplanet systems? How special is Earth within all the possible worlds? What are the necessary conditions for life to emerge and thrive? What is the fate of the Solar System?" and, "Are we alone…?"

A very wide range of ground-based instruments and facilities are prioritised to support this science, covering planet formation, exoplanet discovery and demographics, exoplanet characterisation, and the late-stage evolution of planetary systems, including: VLTI MIDI, MATISSE and GRAVITY+; VLT instruments SPHERE, MUSE, ERIS, HiRISE (SPHERE/CRIRES+ coupling), RISTRETTO, and possibly MAVIS; ALMA; Rubin LSST; ESO CRIRES+, NIRPS. Precision radial velocity measurements are critically important to this field, with the CFHT SPIROU, ESO NIRPS, and the Spanish/German CARMENES explicitly cited in the near-infrared, while in the optical the VLT ESPRESSO, the US EXPRES facility, and HARPS3 on the Isaac Newton Telescope are all explicitly cited. VLT and (later) ELT direct imaging are prioritised. LOFAR and SKA (both with substantial UK interests) are cited for the study of star-planet interactions. The wide-area multi-object spectroscopic surveys WEAVE, DESI, 4MOST and SDSS-V are particularly important for the study of white dwarf planetary systems. Further in future the report prioritises the ELT instruments MICADO, HARMONI, METIS, HIRES (now named ANDES), PCS. Furthermore, laboratory astrochemistry (also a UK area of world-leading expertise) is explicitly prioritised in order to complement and interpret astrophysical observations. Theoretical work with three-dimensional numerical simulations is similarly prioritised.

*2.3.6 Understanding the solar system and conditions for life*
This chapter of the ASTRONET roadmap also covers a wide range of UK research interests, and the UK also participated in the working group that drafted this chapter. The scientific remit is solar system science and heliospheric physics, and as such is much more closely aligned with the specialism of SSAP than AAP. A large proportion of this review is also based on space missions that fall within the UKSA remit. For both reasons we do not summarise this chapter here. Nevertheless, many of the same priority terrestrial facilities and technologies are in common between these solar system recommendations and those of AAP, including Rubin LSST and laboratory astrochemistry.

*2.3.7 Computing - big data, high performance computing, data infrastructure*
This overarching technology chapter of the ASTRONET roadmap covers high performance computing, software engineering, data products, software tools, open science (including astronomical use of the European Open Science Cloud, an interdisciplinary open science initiative of the European Commission funded by approximately a third of a billion euros to date), and the carbon footprint of astronomical computing. UK expertise was represented on the working group that drafted this chapter.

The report provides six main recommendations, which in summary are:
1. Developing and investing in a professional software engineering / computational skills base in Astronomy, including career development with clear progression pathways in academia and improving the diversity of the workforce, and new measures to quantify the impact and usefulness of computationally focussed outputs.
2. Missions and facilities should plan an integrated approach for data products and software tools including design, delivery, maintenance, development, and scientific data preservation.
3. Creation of a "tiered" approach to data infrastructures, for all astrophysical data including models, simulations and mock data.
4. A fully collaborative, open and synergistic approach to the astronomy-computing ecosystem, encompassing data, software, processing, analysing and modelling, and embedding reward structures to realise this ambition.
5. ASTRONET should produce or commission a biennial quantitative report to assess the carbon footprint of computing in Astronomy.
6. ASTRONET should develop specific actions to coordinate cross-cutting activities.

These aspirations are shared by the UK community, but AAP notes these aspirations are not currently being met. For example, they sit uneasily with the AAP consultation responses below that highlight current UK deficiencies in the support of career pathways for technical roles in astronomy. **See Recommendation 3.4 below.**

*2.3.8 Societal aspects – education, public engagement climate action, equality, diversity and inclusion*
At the time of writing (November 2022), this chapter of the ASTRONET roadmap is in draft form and is not yet finalised. The draft report covers the social and cultural relevance of astronomy, education, technology transfer from astronomy, public engagement, astronomy for sustainable development, climate action, gender equality, and inequalities in astronomy. The educational and cultural aspects of the report are not germane to our AAP roadmap so we do not summarise them here, but our AAP consultation responses below share many of the same ideas as the diversity, equity and inclusion themes of this report, including: "significant gender imbalance still exists within astronomy as a profession [...] there is a significant imbalance in the inclusion of underrepresented and vulnerable groups (persons with (dis)abilities and persons with racial, ethnic, religious, LGBTQI+ backgrounds) [...] The European astronomical community needs to develop specific training [for] researchers and advocate for equity and inclusion". Furthermore, the draft report is currently recommending that "European astronomy (community) should (at the very least) follow the European timeline towards carbon-neutrality: 50% reduction of $CO_2$ emissions by 2030 and 100% by 2050"; AAP are pleased to note that the environmental impact was one of the criteria in assessing the 2020 UKRI infrastructure funding bids, and AAP welcomes the 2020 UKRI Environmental Sustainability Strategy[30] that aims to achieve net zero carbon emissions from UKRI no later than 2040.

*2.3.9 Extreme astrophysics*
At the time of writing (November 2022) this chapter of the ASTRONET roadmap is currently in draft form. This science theme again covers areas in which a wide UK community are active and in which the UK has many leadership roles, and UK expertise was represented in the working group that drafted this chapter. Extreme astrophysics here refers to the environments within and around compact objects such as white dwarfs, neutron stars, and black holes. At the time of writing there is some overlap with the science themes of the "Formation and evolution of stars" ASTRONET roadmap chapter (see section 2.3.4 above). Key science questions include: What is the nature of matter at nuclear densities? Where are the heavy elements made? How do compact objects produce energy and accelerate particles at all scales? What is the origin of cosmic rays of all energies? How do compact objects form and evolve? To what precision can general relativity describe gravity? What new fundamental physics can be probed with extreme astrophysical objects?

Many of the key facilities supporting these science questions are space-based and therefore largely outside the remit of the AAP roadmaps, but a wide range of ground-based facilities and instrumentation are also prioritised. Current key ground-based facilities cited include: LIGO/Virgo/KAGRA, neutrino detectors (e.g. the US-led IceCube and the European-led KM3NeT), LOFAR, Pierre Auger, MAGIC, HESS, the South African SKA precursor MeerKAT, JVLA, EVN, ground based 2-8 m telescopes including wide-field optical/IR facilities. Future ground-based facilities that will drive future discoveries include: SKA, CTA, ELT, Rubin, Einstein Telescope, the Italian ASTRI facility, the Global Cosmic Ray Observatory, the Giant Radio Array for Neutrino Detection (GRAND), and the next generation Event Horizon Telescope ngEHT. The neutrino, gravitational wave and cosmic ray capabilities fall within the PAAP remit, but the remainder have a striking commonality with the facilities supporting other astronomy areas in both the ASTRONET roadmap and the AAP consultation responses on community priorities below.

---

[30] See https://www.ukri.org/wp-content/uploads/2020/10/UKRI-050920-SustainabilityStrategy.pdf

## 2.4 Subsequent UK developments and the scope of the AAP roadmaps

Since the AAP consultation closed, UKRI postgraduates wrote an [open letter to UKRI](#) on the impact of the cost of living crisis combined with stipends being calculated on the previous year's inflation, and requesting that stipends are uplifted to match the current inflation rate. AAP are very pleased to see UKRI increase the 2022-3 stipends in response to this community input.

Also since the AAP consultation concluded, STFC launched a further consultation on the changes to the consolidated grant system. This new consultation has just reported its findings, with applications accepted for the new scheme in March 2023.

The present document condenses the call for white papers and the consultation into science and technology roadmaps. However, **AAP have some concerns about potential mis-uses of the roadmaps.** These roadmaps only describe the community's present strengths and aspirations, and an overview of the community's near-term and mid-term priorities for science and technology opportunities. The roadmaps are not suitable for drawing a funding threshold line if the core programme funding is reduced, because that would need either an STFC-wide Balance of Programmes review with an Astronomy Evaluation Panel, or an equivalent but Astronomy-specific tensioning of spending commitments, which in either case would be driven by explicit evaluation criteria and explicit funding envelope options, as well as having agreed expected near-term outcomes on the basis of the recommendations. AAP do not believe it would be helpful to provide a finer-grained long-term prioritisation than is presented in this roadmap given that the criteria, prospective funding envelopes and outcomes are all nebulous, because **(a)** the nebulous constraints leave too great a risk of mis-construing or mis-stating the community views **(b)** the evidence base for community views and quantitative measures of facility outputs/impact and other as-yet-undetermined assessment criteria either do not yet exist or have not yet been assembled, and **(c)** implementation of a finer-grained priority list without considering the wider impacts on the core programme would expose astronomy to the risks of unintended consequences in taking on new commitments or divesting from existing ones. Therefore, the roadmaps must not be used to award funding to something that appears to be a priority, without any discussion with AAP about the consequences for other parts of the programme and any developments since the roadmaps were drafted.

**Recommendation 2.1: The 2022 AAP Science and Technology Roadmaps should not be used in isolation to evidence current community support without further reference to AAP, because astronomy is a fast-paced and rapidly changing field. Nor should these roadmaps be used to draw a funding threshold line in the event of changes to the core programme funding, because the Balance of Programmes Review (or a hypothetical Astronomy-specific near-term tensioning exercise) is the appropriate process for that.**

# 3. Summary of the 2021-2022 AAP Community Consultation

## 3.1 Overview of the process

In this section we summarise the response to the AAP community consultation in November 2021, in which we called for community input on priorities, as well as for science and technology white papers (see below). Many of the themes highlighted in the 2021 community consultation overlap with the earlier AAP consultation in 2018 that fed into the Balance of Programmes review. AAP received only 81 responses compared to 293 in 2018. AAP suspects time pressures related to working from home and the pandemic. There was particularly low representation among early career researchers. In view of this low response rate AAP also solicited further free-form feedback from the community in 2022. The consultation itself is provided in a separate document[31] so we only summarise the findings here.

## *3.2 Overview of white papers*

An important part of developing our science and technology roadmaps was the call for community white papers. These were intended to have the possibility of also serving a dual purpose of assisting STFC with having a 'bank' of potential projects that could be put forward in the event of UKRI-wide ad hoc funding opportunities (see below). AAP are supportive of STFC's aspiration to make the Infrastructure Funding more strategic with more frequent calls to the community, and less dependent on last-minute calls. The STFC Visions process (which replaces the previous Priority Projects calls) is explicitly intended for inter-disciplinary/cross-disciplinary large infrastructure projects across UKRI, and the white papers discussed here did not explicitly require this cross-UKRI remit. At least some of the white papers discussed in our roadmaps may fit the remit of a future STFC Infrastructure Fund.

From our community responses it is apparent that both the white paper and infrastructure funding submissions are fundamentally only a snapshot of the ideas present in the community, so it is not obvious that it is possible to create a long-term roadmap plan that maintains the flexibility and responsiveness to changes in the scientific and technological landscapes.

AAP were impressed at the breadth and depth of ideas submitted by the community. AAP members independently evaluated the relative priorities of the white paper submissions and where possible determined consensus views. Besides the specific projects and facilities discussed in the roadmaps below, AAP received a wide range of science themes in the white paper submissions with clear evidence for UK leadership, including the following.

- What is the chemical and mineralogical composition of astronomical dust? This interdisciplinary white paper covers Galactic planetary and interstellar astronomy, astrochemistry, planetary science and meteoritics, and infrared spectroscopy. The paper makes a compelling case for the support of laboratory astrochemistry and astro-mineralogical research to underpin the interpretation of major data sets from facilities that represent top community priorities, e.g. those from JWST. The white paper covers areas of interest of both AAP and SSAP, and is in concordance with the European ASTRONET priorities (section 2.3.5).
- Euclid Science Exploitation in the UK is a white paper submitted by the Euclid:UK coordination group. There are around 350 UK-based researchers working in the Euclid Consortium. The science themes are extremely wide, covering cosmology, extragalactic astronomy, Milky Way and the local universe, as well as SSAP areas of interest, and are very well aligned with STFC Science Challenges[32] as well as with many of the ASTRONET themes (section 2.3). STFC funding of the science exploitation is within the remit of AGP. A related white paper was received by AAP on the new era of strong gravitational lensing science, covering not just the discovery space opened by Euclid but also that of Rubin LSST, SKA, and the NASA Roman space telescope, with follow-ups facilitated by ground-based 4-metre class facilities (e.g. ESO NTT, 3.6m, WHT, VISTA/4MOST), 8-metre class telescopes (e.g. ESO VLTs) and 30-metre class telescopes (ELTs first light 2027), as well as space telescopes (e.g. JWST). The step change in available discovery space covers a wide range of science goals, including cosmography, dark matter structure and substructure, quasar accretion disc structure and highly magnified background source morphologies and transients. The goals are in concordance with a broad range of ASTRONET aspirations (section 2.3).

---

[31] See https://www.ukri.org/publications/astronomy-advisory-panel-summary-of-2021-community-consultation/
[32] See https://www.ukri.org/publications/stfc-science-challenges/stfc-science-challenges-in-frontier-physics/

- The white paper submitted to AAP on Stellar Variability and an Extreme Precision Radial Velocity Roadmap in the UK presented, in effect, a brief update to the 2015 STFC review of exoplanet science[33]. The UK has clearly established itself as a world leader in this area. The large author list argues for **developing a stellar variability and extreme precision radial velocity roadmap** in liaison with other international bodies such as ESA, ESO, NASA, NSF etc. The facilities driving progress in this area include HARPS-N at the Italian TNG, the Solar Telescope, HARPS at the ESO La Silla 3.6m, CARMENES at the CAHA 3.5m, ESPRESSO at the ESO VLT, NEID at the WIYN 3.5m (a US facility with limited UK access). For characterising solar/stellar variability, the white paper highlights facilities in the SSAP remit including SDO, BiSON, DKIST. Looking forward, the white paper cites the importance of the UK-led HARPS3 at INT, led from the UK, currently under construction (first light expected 2023) and HIRES (now named ANDES) at ELT (under construction) which has passed Phase A.
- The exploration of free-floating planets. This science white paper aims to use VLT, ELT, Rubin LSST, and space missions Roman and JWST. The science theme is an area of UK strength that features only implicitly in the ASTRONET roadmap (section 2.3.4) and may represent a key opportunity for UK international leadership.
- REACH: Radio Experiment for the Analysis of Cosmic Hydrogen[34]. This project aims to detect the signals of cosmic reionization through sky-averaged 21-cm mapping, addressing the instrumental systematics present in other experiments, and stealing a march on the SKA. REACH is a collaboration between the University of Cambridge and Stellenbosch University in South Africa. Although the facility is not currently financially supported by STFC, AAP support the aims of the project and are happy to include it among the high priority facilities below that represent a selection of the current and aspirational portfolio of "small" UK astronomy facilities.
- The role of massive stars in galaxy evolution and nucleosynthesis, supporting facility use of VLT, e-MERLIN, ALMA, ELT, SKA, 4MOST, WEAVE, etc., as well as space missions, and with scientific aspirations in concordance with those of ASTRONET (section 2.3.4).
- Eclipsing binary stars: high-precision probes of the physics of stars and planets. This white paper covers an area of UK strength, and supports the use of WASP, NGTS etc in conjunction with a variety of space missions. The science area is well aligned with ASTRONET aspirations (sections 2.3.4 and 2.3.5).
- Pulsar rotational glitches, a probe to the extreme physics of neutron stars; Precision pulsar timing as a tool for probing fundamental physics; Radio emission from pulsars and the pulsar magnetosphere; Revealing the Galactic neutron star population through pulsar searches. These comprehensive and wide-ranging white papers cover fundamental physics from neutron stars, supporting facility use of e.g. Jodrell Bank, Parkes, MeerKAT/MeerTime, LOFAR/LOFAR2.0, SKA, as well as a variety of space missions. Compute is also a key requirement, particularly also in an additional white paper on magneto-thermal evolution of neutron stars. A further white paper on optical follow-up of binary neutron star systems made a strong case for investigating the dynamics, geometry, and probes of the physical processes, with the aid of ESO NTT, VLT, ELT, as well as GOTO, Rubin LSST, and the Spanish facility GranTeCan. The white papers are in concordance with aspirations in ASTRONET (sections 2.3.4 and 2.3.9).
- Multi-messenger astronomy featured in a wide range of white papers submitted to AAP. The white paper on time-domain polarimetry as a unique discovery space for transients covered the important emerging fields of multi-wavelength, multi-messenger time domain science. Besides aspirations for new instrumentation, the facilities supported by this white paper

---

[33] See https://www.ukri.org/about-us/stfc/planning-strategy-reviews/exoplanets-uk-research-review/
[34] See e.g. de Lera Acedo, E., et al., 2022, Nature Astronomy, 6, 984, https://doi.org/10.1038/s41550-022-01709-9

- include Rubin LSST, the Zwicky Transient Factory, Liverpool Telescope, NRT, VLT, WHT, ALMA, SKA, and a variety of space missions. Multi-messenger astronomy is an emerging priority of both ASTRONET (section 2.3.4) and APPEC[35], and naturally also features in the PAAP roadmap. A further white paper on the high energy transient Universe underlined the importance of X-ray and gamma-ray space telescopes for multi-messenger astronomy (formally outside AAP roadmap remit) and the importance of complementary ground-based observations from e.g. SKA, ELTs, CTA, LOFAR2.0, GOTO; the science goals are broad and cover emission mechanisms of gamma-ray bursts, relativistic jet physics, thermal transients, and the early Universe. AAP also received a white paper on searching for fast radio bursts (FRBs) and other radio transients. This white paper addresses a very topical research area, and also covers the emerging important techniques of multi-wavelength, multi-messenger astronomy (see also the ASTRONET roadmap above, section 2.3.4). The science case covers FRBs as cosmological probes, potential FRB progenitors such as magnetars, slow Galactic transients, and population phenomenology. Key facilities include MeerKAT, Jodrell Bank, e-MERLIN, European VLBI Network (EVN), LOFAR and its 2.0 upgrade, SKA, and a wide range of X-ray space telescopes.
- Gravitational Wave Astronomy: Advanced LIGO+, Cosmic Explorer, LISA, Einstein Telescope, Pulsar Timing Arrays. These science themes also feature in the PAAP and ASTRONET roadmaps (section 2.3) and reflect a widespread interest in this new astronomical window, with UK leadership in both science and technology. AAP also welcomed a white paper on neutron star physics from gravitational waves, supporting also multi-messenger / multi-wavelength observations with LOFAR/LOFAR2.0, Lovell Telescope, and SKA, underlining the overlapping interests of PAAP and AAP.

**Recommendation 3.1: STFC and UKSA should support the UK exoplanet community in developing a stellar variability and extreme precision radial velocity roadmap in liaison with other international bodies such as ESA, ESO, NASA, NSF etc.**

## 3.3 Top, overarching community priority in the consultation

**Exploitation, i.e. grants,** remain the top community priority and are chronically underfunded, despite a very welcome uplift of £2M year on year (leading to a £6M increase by 2024/25) that was announced at the 2022 National Astronomy Meeting from the last Comprehensive Spending Review outcome. This funding for human capacity is now the main limiting factor for our national science capability in astronomy, but it is a UKRI-wide problem. As John Womersley put it six years ago, we are still *"paying for gym membership and being unable to afford the bus fare to get there"*. The continued, chronic underfunding of exploitation has arguably led to the result that 53% of the community are now dissatisfied or very dissatisfied with the balance between exploitation, operations and development; in contrast, our 2018 exercise had 52% wanting the exploitation line to grow only if there were new money. Astronomy's UKRI exploitation funding gap has been partly counterbalanced by EU funding, reflecting also our strong relationships with continental communities and beyond, so our association to the Horizon programmes is critical, providing not just funding but also close association with continental-scale research networks.

Several white papers (see above) mentioned the possibility of hypothecating postdoc/fellowship funding, i.e. setting aside some PDRAs/fellows for particular science areas or to support particular facilities. AAP believe the underlying driver for these submissions is that exploitation funding (i.e. grants) are so chronically under-supported and that the limiting factor for the UK's science research capability in astronomy is human capacity; however, we suspect that allocating such a sparse resource as postdoc/fellowship funding to particular science areas would prove too contentious and divisive to be implementable in practice within the core programme while maintaining community support. If additional UKRI funding were found beyond that of the STFC core programme, then dedicated subject-specific fellowship opportunities may prove less contentious.

---

[35] See e.g. https://indico.ego-gw.it/event/199/

**Recommendation 3.2: STFC should prioritise increasing the exploitation line, which is the community's top priority.**

**Recommendation 3.3: UK association to the EU Horizon funding programme is critical for astronomy. Replacing the funding alone would alleviate the exploitation pressure but would disconnect the UK from continental-sized research networks.**

Our white paper and consultation submissions highlighted the problem of a lack of career structure and career opportunities in the cross-cutting areas of technology development, software, and instrumentation. Universities can find it difficult to recruit and retain core technical teams. The white papers suggested aspirations for early-career fellowships for instrumentation; expanding the number of research software engineer fellowships in the STFC area; helping universities retain core technical teams via grant assessment guidelines to facilitate employing staff on multiple grants. In the Equity, Diversity and Inclusion section of our consultation, it was also highlighted that there is little recognition of the software engineer / blended (software and astronomy) as a career path in an environment, yet this is crucially important (and we may add that a large amount of the effort is done by early-career researchers despite the whole community benefitting). The white papers aspired *"to see a visible shift of priority from the science exploitation of facilities to their delivery in order to highlight instrumentation* [and AAP would add software and project roles] *as a viable career specialism. Such a priority shift would also extend the reach of astronomy-based work to stakeholders in other critically important domains."* We agree and see some of the underlying problems not just being structural but also arising from the chronic underfunding of exploitation. There is no headroom. This is not just limiting our science exploitation now, but is incurring long term, strategic costs. The white paper submissions also correctly pointed out that the typical short-term (e.g. 3 year) project grant durations are not commensurate with typical project lifetimes that are a factor of several longer, creating an avoidable career precarity particularly for those in technical roles that are increasingly important for STFC science exploitation.

**Recommendation 3.4: STFC should review the career structure for instrumentation and technical roles, both within and beyond astronomy and discuss with Universities as to how to implement that within their structures.**

## 3.4 Highest rated science and technology themes/facilities in the consultation

SKA (together with its pathfinders/precursors) and ESO (including ELT) remain highest rated for both the science and technology roadmaps in our consultation. Besides the consultation response, the SKA breadth of community support is also very well evidenced by UK participation in SKA working groups[36] as well as regular town hall meetings[37] organised by the UK SKA Science Committee. AAP are very pleased that the SKA has been awarded UKRI infrastructure funds for the development of the UK SKA Regional Centre. ESO being among the highest community priorities is also in accord with other obvious lines of evidence besides our consultation response, such as the recent evaluation[38] of UK membership of ESO, the breadth and depth of UK PI and co-I time awards[39] on current ESO facilities, the factor ~8 oversubscription[40] on Europe-node ALMA time (in which the UK is typically one of the largest national European winners of PI and co-I time), the breadth of UK involvement[41] in science working groups[42] and instruments[43] on the forthcoming Extremely Large Telescope (ELT), and the community participation in UK ELT town hall meetings[44].

---

[36] See https://astronomers.skatelescope.org/science-working-groups/
[37] See https://www.ukri.org/about-us/stfc/how-we-are-governed/advisory-boards/ukskasc/
[38] See https://www.ukri.org/publications/socio-economic-impact-evaluation-of-the-uk-subscription-to-eso/
[39] See https://www.eso.org/sci/observing/teles-alloc/all.html
[40] See https://almascience.eso.org/documents-and-tools/cycle9/cycle-9-proposal-submission-statistics

**Recommendation 3.5: STFC must maintain the UK role in the SKA and support the development of the UK SKA Regional Centre.**

**Recommendation 3.6: The UK must remain a member of the European Southern Observatory and play leading roles in its development of its world-class instrumentation, including the second- and third-generation instrument suite of the ELT and the development of ALMA instrumentation.**

High performance and high throughput computing provision also remain a community priority. This has received a welcome boost in 2020 in the form of £20m of capital funding from the UKRI World Class Laboratories funding line, enabling the long-awaited deployment of the DiRAC-3 phase 1 upgrade in 2021. The establishment of the IRIS[45] project, following capital grant funding from BEIS in 2018, has also provided a framework for linking the range of digital research infrastructure that falls under the STFC remit. However, large oversubscription factors and the ever increasing data requirements of state-of-the-art simulation codes and large-scale surveys mean that continued support remains vital.

**Recommendation 3.7: UK HPC capabilities such as DiRAC & IRIS underpin a wide range of world-leading UK theoretical astrophysics and data science that must be continually supported and upgraded to remain competitive.**

## 3.5 Emerging technologies or capabilities identified by the community for inclusion in the roadmap

The largest response was the development of improved detectors, in particular energy-sensitive detectors (KIDS, MKIDS). Many replies highlight the importance of existing or upcoming projects. ALMA, LOFAR, SKA, JWST, LSST, GRST are mentioned several times, while Euclid, CTA, new instruments on the VLTs (in particular optical/NIR interferometry), gravitational wave research, exoplanet science, high-energy instruments and improved radio telescopes were also mentioned. Big Data and machine learning techniques were also mentioned as an opportunity.

## 3.6 Science challenges identified by the community as requiring major investments

Many science challenges were identified by the community as requiring major investments, including gravitational waves from space to Gamma-ray detectors, a new all-purpose space observatory, supporting ESO, various space missions, studying planets, discovering life, strengthening staffing for research, HPC and a new sub-millimetre telescope.

## 3.7 Equity, Diversity, Inclusion, and Careers

Science, including astronomy, does not always operate as a meritocracy, and there are many well-recognised biases in accessibility, inclusion and career progression. Improving equality, equity[46], diversity and inclusion (EDI) is therefore part of the core business of astrophysics, besides also having a moral imperative. The consultation presented a wide range of community suggestions. **See also Recommendations 3.4 and 4.2.** Some suggestions were within the remit of the Consolidated Grant consultation process, which has now completed. Other suggestions that extend across UKRI and beyond the remit of AAP included:

- The career structure disadvantages some and negatively affects EDI objectives. For example: differences in leave (including maternity) for professional staff, and no maternity leave on grants; early career precarity, especially for those with low-income backgrounds; deficits in support for disadvantaged/under-represented communities; the need to move overseas; the need to improve leave systems for those with caring responsibilities, and lack of support for families to attend conferences; ableism and the ability to travel; work-life balance; unsafe/unhealthy working environments, i.e. not free of prejudice/abuse; insufficiently diverse set of role models at all levels.
- EDI progress evidence should be collected for both institutions and individuals, with demonstrable outcomes to avoid tick boxes. EDI training should be mandatory and action should be taken against staff who disregard EDI, while good EDI practices should be recognised and rewarded. Why do we weight science track records highly, and not even ask for EDI track records?
- Research consortia should be expected to implement and adhere to their codes of conduct.
- STFC should monitor and report on the diversity of its own advisory structures.

---

[41] See https://www.elt-uk.org/astronomers/
[42] See https://elt.eso.org/science/
[43] See https://elt.eso.org/instrument/
[44] See https://www.elt-uk.org/astronomers/meetings/
[45] See https://www.iris.ac.uk/
[46] See e.g. https://social-change.co.uk/blog/2019-03-29-equality-and-equity

# 4. UKRI Infrastructure Funding call: AAP and community responses

Subsequent to the consultation and white papers call, AAP solicited community input for the UKRI Infrastructure funding "Preliminary Activity Wave 3" call in Summer 2022. Five bids were received, of which AAP were only allowed to recommend three due to the proposal demand management:

1. A 10-12-metre spectroscopic survey telescope. This design study aimed to prepare the way for a very wide range of science including galactic archaeology, galaxy assembly and the cosmic web, and transient science. This is a natural next step in Galactic and extragalactic survey parameter space: the next generation imaging surveys by the Vera Rubin Observatory and ESA Euclid will create targets that are too faint for 4-metre class observatories, and both the number of targets and the near-all-sky coverage make follow-ups by current 8-10-metre class observatories unfeasible. The project covered both the Mauna Kea Spectroscopic Explorer and the proposed ESO Widefield Spectroscopic Telescope. This wide-field spectroscopic capability has already been identified as a European community priority by ASTRONET (see above), and additionally featured in the AAP call for white papers.
2. UKELT: The next generation instrumentation suite for the Extremely Large Telescope. This design study aimed to prepare the way for the second generation of ELT instruments: the multi-object spectrograph MOSAIC, the high-resolution spectrograph ANDES, and the planetary camera and spectrograph PCS. The UK is in a leading position for ELT instrumentation, which in turn will be transformative across essentially all of UK astrophysics. This study had previously had some of the phase A & PCS research and development staffing costs funded by PPRP but not capital or hardware costs for the instruments. Unlike previous VLT and first-generation ELT instruments, ESO will not provide the hardware costs for second & third generation ELT instruments leaving the responsibility of raising hardware funds to the instrument consortia. The project also featured in the AAP call for white papers, further evidencing UK community interest.
3. Scalable continuum cameras for sub-mm telescopes. This design study aimed to prepare for a

first-generation instrument for the AtLAST large submillimetre-wave telescope, which in turn has had a facility design study funded by the EU Horizon 2020 programme. In the process, this project would also deliver a wide-field continuum camera for the JCMT, giving the UK a tactical and strategic advantage for feeder programmes for ALMA across a wide range of astronomy, including extragalactic surveys, Galactic star formation and protoplanetary discs. The UK JCMT involvement has been jointly funded by PPRP and university contributions, while AtLAST is a clear European priority across a wide range of science goals (see above) and featured in the AAP call for white papers, further evidencing UK community interest.
4. UK Extreme High Frequency Facility. This project aimed to create a technological development hub for coherent detector development from radio to sub-millimetre wavelengths. The project would support UK instrumentation interests for SKA, ALMA and several other facilities, and would cover a wide range of science interests from star formation to extragalactic surveys.
5. Joint Rubin–Euclid data processing. The ground-based Rubin Legacy Survey of Space and Time will image the whole available sky twice a week for ten years at optical wavelengths, while the ESA Euclid mission will map around half the sky at higher resolution at optical wavelengths, extend to near-infrared wavelengths and provide near-infrared spectroscopy. The Rubin and Euclid consortia together have comprehensively studied[47] the wide-ranging science cases resulting from combining these data sets, and this project proposed to create new value-added data products from the joint analysis. The astronomy science goals span exoplanets, Milky Way, nearby galaxies, transient phenomena, galaxy formation and evolution, and cosmology.

All five were excellent, and well-aligned with our roadmaps below. Numbers 1-3 (in no priority order) were selected. (STFC then down-select all the proposals across all the advisory panels to a total of just three to submit to UKRI.) We were advised by UKRI that our number 5 might not fit the remit of that particular call, and number 4 was deferred to the next call based only on grounds of proposal readiness at the time of the deadline. These submissions did however unveil two facets of the UK community: firstly, the independent, internationally-excellent groups working on sub-mm/mm-wave science and technology do not yet have a single consistent voice (see also section 5.5.2); secondly, the difficulty in finding funding routes for joint facility data processing reveals a deficiency in support for digital infrastructure, despite big data and machine learning being highlighted by the UK astronomy community as important emerging technologies (see above).

**Recommendation 4.1: STFC (via AAP) should commission a review of UK submm/mm-wave science and technology, covering UK aspirations for current and future large single-dish facilities that feed the major international interferometers, and the underpinning aspirations for next-generation instrumentation, identifying areas of international excellence.**

**Recommendation 4.2: There should be increased support for digital infrastructure in UK astronomy, including the creation of value-added data and software products, open science, and the development and implementation of machine learning technologies. See also Recommendation 3.4.**

# 5. Science and Technology Roadmap

## 5.1 Overview
In constructing this roadmap, we refer both to large overarching projects (e.g. ESO, LSST, SKA) and specific applications (e.g. joint LSST/Euclid data processing). As shown comprehensively above in section 2, every facility supporting astronomy typically covers a very wide range of science themes, and the science themes themselves demand a diverse range of facilities and instrumental capabilities. This is driven ultimately by the diversity of physical processes that occur in each science theme, and the common ground-based technological approaches to investigating these processes throughout the science themes (e.g. far-infrared to millimetre wavelengths for molecular astrophysics and dust; radio astronomy for synchrotron and coherent processes and atomic gas; UV, optical and near-infrared astronomy for ions, stellar populations and black hole accretion). Major discoveries in observational astronomy are often driven by the opening of new observational parameter spaces and/or developments in instrumentation. Therefore we have opted to combine the science and technology roadmaps in a single roadmap for technology and facility needs and aspirations.

---
[47] See https://arxiv.org/abs/2201.03862

State-of-the-art instrumentation and facilities drive developments in areas throughout the STFC "Frontier Physics" science challenges. There is a wealth and depth of key technologies where the UK is a leading force, including:

- Kinetic inductance detectors used for a range of aspects such as spectroscopy, imaging, time-domain, CMB, galaxy evolution, etc.
- Software, computing, receivers, and development for radio astronomy (Jodrell Bank Observatory, e-MERLIN, Lovell Telescope, Square Kilometer Array).
- Infrared detectors
- High performance computing

AAP would add (astro)photonic technology to these community-driven suggestions. Furthermore, many technologies have the capacity for cross-cutting or multi-disciplinary applications, such as: areas that require mathematics and quantum devices development, time-domain astronomy, machine learning, high performing computing, artificial intelligence, data science/big data, and communications. Submm astronomy technology was identified by the community as having a wide range of high priority science exploitation areas. There is a very wide range of opportunities for the UK to play a leading role in large international astronomy projects, covering all wavelengths of observational astronomy, including especially the Vera C. Rubin Observatory, SKA, Euclid, LISA, and ELT. International leadership opportunities also exist in theoretical astrophysics, reflecting our high priority technology item below. Although outside the STFC remit, there is also a very strong community appetite for UK international leadership opportunities via space agency bilaterals (e.g. NASA/China/Japan).

There is currently a gap in funding resources for exploratory instrumentally-focused work, compared to that available for science exploitation and technology development with clear near-term goals. There are also structural problems in supporting the career development of instrumentation specialists.

Some important technology roadmap conclusions were suggested directly by the community, and which AAP endorses:

- The strategic importance of trading unique UK technical capabilities, often mission-critical, for a seat at the table setting the scientific agenda of a mission.
- "Creating consistent opportunity for scientific leadership of future missions will require a strategic plan to link university-led innovation to the longer-term resources needed to support future major technical leadership bids."
- "We do not believe anyone can unambiguously identify the technologies needed to address STFC's key science challenges in the next decade. So to remain competitive there must be strong UK investment in a broad program of advanced instrumentation and its supporting technologies" which the authors suggested should include precision optomechanics, nm-resolution metrology and both electromagnetic and non-electromagnetic sensor development.

**Recommendation 5.1: There must be strong UK investment in a broad programme of advanced instrumentation and its supporting technologies, including (but not limited to) precision optomechanics, (astro)photonic technology, nm-resolution metrology, sensor development, kinetic inductance detectors and infrared detectors in general, CCDs and CMOS detectors, receiver development, low noise high-electron-mobility transistors, device fabrication capabilities, software and computing.**

AAP is pleased to note that there is now a dedicated STFC early TRL funding stream with follow-on fund, which had its first round in 2022. This covers all the STFC core science programme but many astronomy groups have applied[48]. The call was associated with a capital call this year for lab equipment etc. (over £1M awarded) AAP are also pleased that STFC expect[49] the second joint UKSA-STFC early TRL Technology Fund for Space Science will shortly be announced offering up to £1.6M for targeted projects in detectors, miniaturisation and robotics for science instruments, and that it is expected this will be repeated in 2023 and 2024 with different themes.

The priorities listed below are for near- and mid-term timescales. Our consultations also asked the community about the anticipated science themes on longer timescales. Community-identified science themes to be addressed in the next two decades include "life elsewhere" and "exoplanet formation and characterisation", studies the nature and composition of Dark Matter and Dark Energy in order to disentangle the processes involved in galaxy formation and evolution, high precision gravitational tests of GR, gravitational waves, including different sources as well as waveform models ahead of LISA.

## *5.2 Top priority science facilities*
The top priority science items have wide-ranging and transformative science goals and world-leading technology development, with strong and very wide-ranging / almost-universal community support.

### *5.2.1 European Southern Observatory*
ESO (including ELT, ALMA) is a clear joint top priority. It has very clear evidence of widespread community support both for near-term science (e.g. VLTs, ALMA) and mid-term (e.g. ELT first and second generation instrumentation) with transformative science. See section 3.4, and **see Recommendation 3.6 above.** The UK Programme for the Extremely Large Telescope positions a wide portion of the UK community strategically in future decades for major science, including in the following science and instrumentation areas:
- Towards Earth: Direct Detection of Exoplanets with the Extremely Large Telescope Planetary Camera and Spectrograph ELT-PCS;
- HARMONI: the first light integral field spectrograph for the Extremely Large Telescope
- ANDES : the High Resolution Spectrograph for the Extremely Large Telescope MOSAIC: the multi-object spectrograph for the Extremely Large Telescope
- METIS: The mid-IR imager and spectrograph for the Extremely Large Telescope

The latter four instruments together cover broad science cases ranging from exoplanets to the epoch of reionization.

### *5.2.2 Square Kilometer Array*
The SKA has very clear evidence of wide community support (including beyond 'typical' radio astronomy groups) with field-changing science across a very wide range of astronomy. In 2022, SKA Observatory was awarded £66.7 million from the Infrastructure Fund in total including future funding years (following endorsement by AAP for funding via the "priority projects" process). See section 3.4, and **see Recommendation 3.5 above.**

---
[48] Colin Vincent, priv. comm., 8 November 2022
[49] Colin Vincent, priv. comm., 8 November 2022

*5.2.3 European Space Agency and other space opportunities*
The ESA Science mandatory programme is formally outside the STFC remit, except e.g. in the science exploitation of current or past missions in AGP, and occasionally in preparatory technology or software infrastructure development for science in PPRP. Nevertheless it covers projects supporting almost the entirety of the UK astronomy science and technology communities, with missions comparable (or greater) in cost, impact and community support to our highest-rated ground-based projects, e.g. JWST, Athena, Ariel, Plato, Euclid, LISA etc. As part of the "dual key" mechanism, AAP advises STFC Science Board when UKSA requests the view of the STFC science community on prospective astronomy missions. Our community consultations highlighted the community importance of several major current and future space missions, despite them being in the UKSA remit. The community support is also well-evidenced by the extensive UK signatories to the respective mission study papers. Rather than consider these individually as with STFC-supported ground-based projects, we have treated these collectively as a single item. At the time of writing there are a number of external policy instabilities, and AAP would not want to send an inadvertent signal by omitting this comparison.

There is a very wide astronomy community appetite for exploring **bilateral space agreements** and other space opportunities. AAP received many excellent white paper ideas across the breadth of the astronomy research portfolio, including:
- The Next Generation of Space missions and their role to understand the most extreme objects in our Universe: Black holes and Neutron Stars
- Dark Ages and First Light: A case for space-based radio cosmology (a science case for participation in, or UK leadership of, a range of potential future space-based 21cm cosmology experiments)
- The search for living worlds and the connection to our cosmic origins (a wide-ranging science case supporting UK roles in exoplanet space missions in the NASA Decadal Review, particularly a large optical/ultraviolet/infrared space telescope)
- Gravitational Wave Astronomy with the Laser Interferometer Space Antenna (also within the remit of PAAP)
- The crucial missing piece in the multi-wavelength jigsaw of the Universe: the Far-Infrared; Participation in a NASA Far Infrared Probe Mission. These white papers have wide-ranging support, high science return, with strategic involvement in mission launching in over 10 years.
- GaiaNIR: Near-Infrared Astrometry Revealing the Galactic Ecosystem
- CASTOR: A Wide-Field, UV Space Telescope
- UK opportunities for extrasolar research with the Twinkle space mission
- Many white papers had extensive aspirations for use of X-ray space facilities.

These opportunities fall outside the STFC remit, but should UKSA seek the views of STFC Science Board on the alignment of these projects with roadmap strategies, then there is already prima facie evidence in their support.

AAP notes that there is currently some discussion of removing all proprietary time from NASA federally funded missions[50,]. US PIs with NASA facility time allocations are often also awarded exploitation grant funding; in contrast, UK PIs have no such guarantee of post-doctoral support to exploit data in these highly competitive allocations. The pressure on UK exploitation grants would further disadvantage the UK community if this proposed policy is enacted (as well as having other consequences).

---

[50] "On 25 August, the White House's Office of Science and Technology Policy ordered departments and agencies to move toward making the results of all federally funded research freely and immediately available by 2026." Clery, D.., 2022, Science, Volume 378, Issue 6619.
https://www.science.org/content/article/should-webb-telescope-s-data-be-open-all

AAP also received a white paper submission on best practices for public engagement with UK space missions. Recommendations included providing sustained engagement support pre- and post-launch, a better awareness of equity, diversity and inclusion in UKSA resource allocation; better recognition of outreach in career progression; retention of hardware created during mission development, for museums; a central resource hub for STEM outreach; and funding for better coordination. These are beyond the AAP remit but AAP encourages the authors to publish their recommendations and/or submit them to the UKSA advisory structures.

**Recommendation 5.2: The ESA mandatory science programme covers projects supporting almost the entirety of the UK astronomy science and technology communities, with missions comparable (or greater) in cost, impact and community support to our highest-rated ground-based projects. The UK must continue ESA membership, and there is widespread astronomy community appetite for exploring bilateral space agreements and other space opportunities.**

## *5.3 Very high priority science and technology*

The very high priority science items have wide-ranging and important science goals and world-leading technology development, with strong and wide-ranging community support. AAP received a community submission in favour of Gravitational Wave Astronomy with Advanced LIGO+ and Cosmic Explorer, and note that this very high priority area is already considered in the PAAP roadmap.

### *5.3.1 High Performance Computing*
**High Performance Computing**, in particular the DiRAC HPC Facility (including capital equipment, operations, training & innovation) has clear UK leadership in many areas underpinning a wide range of UK astronomy. **See Recommendation 3.7 above.**

### *5.3.2 Vera Rubin Observatory*
The Vera Rubin Observatory Legacy Survey of Space and Time covers a very wide range of high priority UK science themes (section 2) and is critical to UK leadership in many anticipated future developments. The major role the UK is playing in the LSST will enable membership for a large fraction of the UK astronomical community; we anticipate Rubin LSST soon achieving the almost-universal breadth of support to rank alongside SKA and ESO. We note above in section 4 that there is a deficit in support for digital infrastructure, such as in supporting the creation of value-added data products from the scientific synergy with Rubin and Euclid; **see Recommendation 4.2 above.**

### *5.3.3 Simons Observatory / CMB science*
Science from observations of the Cosmic Microwave Background are an obvious very high priority with long-term historic UK leadership internationally. The large UK CMB community has a well thought-out strategy with Simons Observatory (SO) as top mid-term priority. SO was not in the core programme of the second BoP review due in part to lack of headroom for major new commitments, but was endorsed by AAP for *new* funding under the "priority projects"; SO was awarded £12.6M of new infrastructure funding from UKRI in 2022 over three years. The CMB community white paper submission to AAP also describes ground-based aspirations beyond SO for CMB Stage 4 and the UK-led CBASS experiment. Further details on CMB science priorities and further smaller facility aspirations are covered in the PAAP roadmap.

## *5.4 High priority science and facilities*
These are science areas and facilities that may not clearly satisfy all of the criteria for the "Very High Priority" areas above, but include projects that could be developed to very high priority, including those with high science return over the coming 10 years. The science areas and projects listed in this section cover a very wide range of costs, science areas and timescales, so we felt it would not be useful to draw prioritisations among them under some hypothetical constraints. Rather, the Balance of Programmes process (or an equivalent Astronomy-specific near-term tensioning exercise) should be used in the light of real constraints. **See Recommendation 2.1 above.**

*5.4.1 WEAVE*
The WEAVE instrument is the next generation spectroscopy facility for the William Herschel Telescope (WHT), built by an international STFC-led consortium. The science goals are wide-ranging, covering Galactic archaeology, white dwarfs, neutral Hydrogen selected galaxies, galaxy clusters, low-frequency radio sources from LOFAR, and quasars. WEAVE has also been highlighted in several science themes of the ASTRONET roadmap (section 2.3 above).

*5.4.2 The Maunakea Spectroscopic Explorer and other 10-12m-class massively multiplexed spectroscopy facilities*
MSE, or the Maunakea Spectroscopic Explorer, is a 11.25-metre optical spectroscopy facility proposed to replace the Canada-France-Hawaii Telescope (CFHT) and sit at the location of the CFHT in an enclosure only slightly larger than the existing dome. The science case is wide-ranging, and spectroscopic facilities of this class are essential for the science exploitation of the coming generation of very wide-area optical and near-infrared imaging surveys from Rubin LSST and ESA Euclid. More details can be found in section 4. AAP supported MSE as a priority for potential UKRI infrastructure funding, and spectroscopy facilities of this class have also been highlighted in several science themes of the ASTRONET roadmap (section 2.3 above).

*5.4.3 LOFAR, the Low Frequency Array*
LOFAR is a new-generation radio telescope operating at low radio frequencies of ~15 to 240 MHz. LOFAR is a precursor facility to the SKA, which is one of the community's top priorities (see above). The science areas covered by LOFAR are vast, from Solar System science to cosmological scales. 23 UK universities are participating in the UK LOFAR consortium. LOFAR is currently operational and highly productive, with UK leadership roles and strong UK science exploitation. LOFAR's staged upgrades, known as LOFAR2.0, are already in progress and the initial stages have approved STFC funding; this will maintain LOFAR's position as the most powerful very-low-frequency and long-baseline radio interferometer internationally until at least 2030. LOFAR features in several science themes of the ASTRONET roadmap (section 2.3 above).

*5.4.4 VLT BlueMUSE*
BlueMUSE is a panoramic 3-D spectroscopy instrument in the UV/blue for the ESO VLT, addressing a very wide range of science from the Solar System to Galactic astronomy to redshifts of z~6. BlueMUSE has been approved by ESO for a Phase A start in January 2023. The consortium is led from Lyon, with the UK bidding for a key role in assembling, aligning and testing the spectrographs. BlueMUSE features in several science themes of the ASTRONET roadmap (section 2.3 above).

*5.4.5 VLT CUBES*
CUBES is a near-ultraviolet spectrograph being built for the ESO VLTs, offering a much higher throughput and sensitivity to the current state-of-the-art VLT UVES instrument (under 5 percent throughput including telescope, atmosphere and instrument). This wavelength range is rich in diagnostic information covering a wide range of science goals covering Solar System science, Galactic and extragalactic astronomy. The project passed its Phase A study review with ESO in June 2021 and started construction in early 2022, with operations planned to start in 2028. CUBES is led by Italy with the UK leading the Science and Detector work packages. CUBES also features in several science themes of the ASTRONET roadmap (section 2.3 above).

*5.4.6 UK National Radio Astronomical Observatory UKNRAO*
UKNRAO is a bottom-up proposal to create a national radio observatory that would be a S&T/R&D coordinating centre for radio astronomy, a timely initiative designed to fully exploit the UK's £200M+ investment in the SKA. This white paper came about partly in response to previous AAP feedback on the diversity of radio proposals in the "priority projects" process, and AAP are very pleased to see that the radio community has engaged in careful, consultative strategic planning to create this white paper. UKNRAO would be established as a Company Limited by Guarantee with a governance structure similar to the highly successful Alan Turing Institute, bringing together all the major players in UK radio astronomy. The science goals addressed by radio astronomy are very broad, as seen e.g. in section 2, and the UK has a large and vibrant radio community. There are excellent synergies with the UK ALMA Regional Centre co-located at one of the UKNRAO sites. The aspirations include a UK SKA Regional Centre, recently funded by UKRI infrastructure funding and perfectly aligned with one of the community top priorities (see above). As well as enhanced national and international standing, UKNRAO would also host key UK instrument and technology development, including a flagship "Receiver Factory" and the UK Extreme High Frequency Facility that was submitted to AAP for the most recent UKRI infrastructure call (section 4).

*5.4.7 Submm Astronomy on Large Scales / AtLAST*
Two white papers made a strong case to AAP for wide-field mapping at sub-millimetre and millimetre wavelengths, to provide the UK with tactical and strategic advantages in creating feeder programmes for the heavily competitive time allocations on the ~£1.3billion ALMA facility. The broad science goals cover star formation, the interstellar medium, galaxies and cosmology. More than 90% of the energy output from newly formed stars and galaxies is hidden by dust and only accessible to these facilities. ALMA excels at deep pointed observations, but even the largest ALMA programmes cover only a few square arcminutes, so wide-field mapping is essential to feed ALMA with targets. The largest submm telescope in the world is the James Clerk Maxwell Telescope (JCMT, with UK roles funded by PPRP and university contributions) and is likely to remain the largest for the next decade, until the AtLAST facility is available; the UK also has collaborative access to the Mexican Large Millimetre Telescope. The UK has a very strong track record in submm/mm-wave astronomy, a broad and vibrant community, and many leadership roles in JCMT Large Programs. The UK also has international leadership in related detector technologies, such as KIDS. Wide-field (sub)-millimetre mapping also features in several science areas of the ASTRONET roadmap (section 2.3 above). AAP recommended the AtLAST instrument feasibility study and JCMT instrumentation as a priority for UKRI Infrastructure funds (section 4).

*5.4.8 Magdalena Ridge Observatory Interferometer, MROI*
The Magdalena Ridge Observatory Interferometer MROI has a broad, transformative science application, from Galactic star and planet formation, to precision stellar astrophysics, to AGN formation. This white paper aspires for ten 1.4m diameter telescopes at MROI; with seven or more telescopes the scientific capabilities of the MROI will far exceed the capabilities of both the state-of-the-art facilities VLTI and CHARA, e.g. double the angular resolution of VLTI and triple the imaging field of view of CHARA. The limiting sensitivity would be 2-5 magnitudes fainter than the current state-of-the-art. The first three telescopes of MROI are currently funded by the US Air Force. The expansion to 10 telescopes was one of the priorities for optical/IR interferometry in the 2021 US Decadal Review. The required technology development would position the UK scientifically **(see Recommendation 5.1)** and establish UK expertise for the Planet Formation Imager (timescales: after ELT first light) and ELT first-generation instruments.

*5.4.9 HARPS-3*
HARPS-3 is a new generation high resolution, high stability echelle fibre spectrograph, supporting the UK's world-leading roles[51] in exoplanet science. HARPS-3 is to be installed on the Isaac Newton Telescope, and aims to conduct precise Doppler spectroscopy of exoplanet systems for the discovery of Earth analogues. The UK-led instrument is part-funded by PPRP and will achieve first light in 2023, and full operations are planned to start in 2024. Half of the observing time will be available to the Isaac Newton Group community. The instrument also features as an exoplanet priority in the ASTRONET roadmap (section 2.3.5).

*5.4.10 Other facility capabilities*
The above high priority facilities include the projects for which dedicated white papers were submitted in our consultation, but there are several others where comparable community support, breadth of science goals and high science impact may also be evidenced. Examples include: UK involvements in DESI, CCAT, 4MOST; VLT/MOONS (six-country collaboration, managed by STFC at UK ATC); GOTO (prioritised by AAP in the previous "priority projects" process and with UK involvement now funded by PPRP); CTA (UK technical interests in instrumentation and science interests in multi-messenger astronomy); laboratory astrochemistry (UK international science leadership underpinning a wide range of science themes); crowd-sourced astrophysics (UK has taken leadership through e.g. the Zooniverse platform). Funding only the top priority facilities is not sufficient for the UK to achieve a scientific return on the investment; the "small" facilities are an essential part of the UK astronomy programme. A facility or instrument being "small" in cost terms is not inconsistent with it being world-leading.

**Recommendation 5.3: Maintaining UK access to the top priority facilities is not sufficient to ensure returns on these major investments, so STFC should maintain and develop a broad portfolio of high/very high priority science facilities, including (but not limited to) WEAVE, LOFAR, JCMT, GOTO, LT/NRT, WASP, NGTS, DIRAC, IRIS, e-MERLIN, Lovell Telescope, among current facilities, and UK roles in e.g. DESI, CCAT, 4MOST, BlackGEM, as well as fostering roles in currently non-STFC-supported projects such as REACH; looking further ahead, HARPS-3, Rubin LSST, Simons Observatory, AtLAST, MSE, CTA, VLT/MOONS. As part of this broad portfolio supporting the highest priority science, AAP further recommends support for laboratory astrochemistry.**

## 5.5 Emerging community priorities
AAP received a number of white papers covering important science and/or technology developments with a significant degree of community support, including the following.

*5.5.1 Astronomy and Space Domain Awareness*
This white paper highlights the importance of space situational awareness, i.e. the study of satellites and debris in Earth orbit. In the UK alone, satellite services underpin[52] £360 billion of economic activity per year. ESA estimates that around one million objects larger than 1 cm are in orbit, all of which are capable of damaging a satellite. Through the STFC 21st Century Challenge Network+, STFC have funded the Global Network on Sustainability in Space, GNOSIS[53], which has successfully facilitated interdisciplinary approaches to the problems of space domain awareness. More broadly, there is arguably good evidence that the 21st Century Challenges have been very effective for fostering societal impact of STFC science, but this is beyond the scope of the AAP roadmaps.

---

[51] See e.g. https://www.nobelprize.org/prizes/physics/2019/summary/
[52] See e.g. https://publications.parliament.uk/pa/cm5803/cmselect/cmsctech/100/report.html
[53] See https://gnosisnetwork.org/

*5.5.2 Line intensity mapping at millimetre/sub-millimetre wavelengths*
The white paper on Line Intensity Mapping at millimetre/sub-millimetre wavelengths offers a roadmap for a new and complementary probe of early universe cosmology. The science goals cover inflation and primordial non-Gaussianity, dark energy through baryonic acoustic oscillations, the growth of large scale structure, and fundamental physics (neutrinos and light relics). The technique also features in the ASTRONET roadmap (see section 2.3 above). While it does not currently have the breadth of science goals and UK community support of the high priority facilities and facility aspirations above, the science is manifestly internationally excellent. A bespoke instrument for the JCMT (where UK access is currently funded by PPRP and university contributions) would provide UK leadership in this area. To support the development of an evidence base for wider community support, AAP recommends that this is considered as part of the wider portfolio UK science and technology in (sub)-millimetre astronomy; **see Recommendation 4.1 above.**

*5.5.3 Development in Africa with Radio Astronomy, DARA*
The Development in Africa with Radio Astronomy programme, DARA[54] is a radio astronomy training project funded by international development funding, through the Global Challenges Research Fund and the Newton Fund, as well as securing matching funding from South Africa's NRF. DARA has also been supported by the EU Jumping JIVE project. Since 2015 it has been delivering basic training in radio astronomy and associated technologies to young people across the eight DAC-list countries partnering with South Africa in hosting SKA-MID. A sister programme, DARA Big Data, has provided MSc and PhD-level advanced training in Africa and in UK universities. AAP commends the creativity and resourcefulness of the DARA team in using astronomy training to address societal challenges in DAC-list countries, but also notes the anticipated cuts to UK ODA funding that will severely curtail these activities. The white paper asks STFC to continue its strategic partnership[55] with NRF in training and placements of researchers and technicians, to which AAP is happy to concur; the five-year agreement started in 2021.

*5.5.4 Collaboration with Goonhilly Earth Station Ltd on Radio Astronomy and Space Applications*
This white paper was submitted on behalf of the CUGAS consortium, the Consortium of Universities for Goonhilly Astronomy and Space. The science goals covered in this collaboration include radio astronomy, interferometry, satellite communications, and space situational awareness. The project seeks to integrate the Goonhilly dishes into e-MERLIN, providing much needed long baselines and higher angular resolutions, while state-of-the-art low noise instrumentation benefits commercial satellite communication roles. Discussions are underway on providing a sovereign UK space situational awareness capability. Further funds are needed, either from CUGAS members or from STFC, for the cost of the data link and hydrogen maser time standard. AAP commend the CUGAS consortium for their creativity and resourcefulness in pursuing this industry engagement.

*5.5.5 High-risk, high return science areas*
AAP were very pleased to receive a white paper on searching for techno-signatures via anomalies in astronomical data, and are happy to record the community support for this area. A proportion of resources must always be spent on high-risk, high-return science.

## 5.7 Projects with a moderate or low science return
No projects submitted to AAP fall into this category, whether through the consultation or white papers or the call for UKRI infrastructure projects.

---

[54] See www.dara-project.org
[55] See https://www.ukri.org/news/forging-a-new-partnership-with-south-africa/

# 6. Scope for further community consultation

AAP have considered whether there is scope and/or community capacity for further major consultations at this stage, following the model e.g. of the PPAP review of the entirety of UK particle physics. With the ASTRONET review still in the process of reporting the last of its findings on the science roadmap for European astronomy (section 2.3), and with the US Decadal Review recently completed (section 2.2), there does not appear to be an obviously compelling case for a UK-specific review of all astronomy, because it would duplicate at least some of this recent and ongoing effort. Furthermore, the reduced survey return rate in the AAP consultation may indicate a decreased UK astronomy community capacity or appetite for further consultation activities, though the global health crisis may also have been a factor. Nevertheless, there are some focussed areas where further consultations may be useful, such as an overview of community aspirations and priorities in sub-mm/mm-wave science and technology **(Recommendation 4.1)**, and an STFC review of the career structure of instrumentation and technical roles both within and beyond astronomy **(Recommendation 3.4)**. The UK CMB community have already organised their own bottom-up community roadmap, updated in the CMB community white paper submitted to AAP and also addressed in the PAAP roadmaps. The UK exoplanet community last had an STFC community consultation[56] in 2015, and our white paper submissions clearly evidence a desire of the UK exoplanet community to develop a stellar variability and extreme precision radial velocity roadmap in liaison with other international bodies such as ESA, ESO, NASA, NSF etc. **(Recommendation 3.1)**; such a roadmap would cover areas of interest of both AAP and SSAP, as well as UKSA. Radio astronomy was last reviewed[57] by STFC in 2017 following the 2016 STFC Balance of Programmes Review, and among several conclusions recommended a review of e-MERLIN from 2022; the radio community has since created a vision for a UKNRAO (section 5.4.6) so the landscape has changed since the 2017 review. Furthermore, the JIVE ERIC, linked to the pursuance of e-MERLIN, is currently three years into its five-year window; e-MERLIN is nominally operational at least until SKA is operational and the role for other radio contributions can be reviewed.

# 7. Summary list of recommendations

**Recommendation 2.1: The 2022 AAP Science and Technology Roadmaps should not be used in isolation to evidence current community support without further reference to AAP, because astronomy is a fast-paced and rapidly changing field. Nor should these roadmaps be used to draw a funding threshold line in the event of changes to the core programme funding, because the Balance of Programmes Review (or a hypothetical Astronomy-specific near-term tensioning exercise) is the appropriate process for that.**

**Recommendation 3.1: STFC and UKSA should support the UK exoplanet community in developing a stellar variability and extreme precision radial velocity roadmap in liaison with other international bodies such as ESA, ESO, NASA, NSF etc.**

**Recommendation 3.2: STFC should prioritise increasing the exploitation line, which is the community's top priority.**

---

[56] See https://www.ukri.org/about-us/stfc/planning-strategy-reviews/exoplanets-uk-research-review/
[57] See https://www.ukri.org/about-us/stfc/planning-strategy-reviews/2017-uk-radio-astronomy-strategic-review/

**Recommendation 3.3:** UK association to the EU Horizon funding programme is critical for astronomy. Replacing the funding alone would alleviate the exploitation pressure but would disconnect the UK from continental-sized research networks.

**Recommendation 3.4:** STFC should review the career structure for instrumentation and technical roles, both within and beyond astronomy and discuss with Universities as to how to implement that within their structures.

**Recommendation 3.5:** STFC must maintain the UK role in the SKA and support the development of the UK SKA Regional Centre.

**Recommendation 3.6:** The UK must remain a member of the European Southern Observatory and play leading roles in its development of its world-class instrumentation, including the second- and third-generation instrument suite of the ELT and the development of ALMA instrumentation.

**Recommendation 3.7:** UK HPC capabilities such as DiRAC & IRIS underpin a wide range of world-leading UK theoretical astrophysics and data science that must be continually supported and upgraded to remain competitive.

**Recommendation 4.1:** STFC (via AAP) should commission a review of UK submm/mm-wave science and technology, covering UK aspirations for current and future large single-dish facilities that feed the major international interferometers, and the underpinning aspirations for next-generation instrumentation, identifying areas of international excellence.

**Recommendation 4.2:** There should be increased support for digital infrastructure in UK astronomy, including the creation of value-added data and software products, open science, and the development and implementation of machine learning technologies. See also Recommendation 3.4.

**Recommendation 5.1:** There must be strong UK investment in a broad programme of advanced instrumentation and its supporting technologies, including (but not limited to) precision optomechanics, (astro)photonic technology, nm-resolution metrology, sensor development, kinetic inductance detectors and infrared detectors in general, CCDs and CMOS detectors, receiver development, low noise high-electron-mobility transistors, device fabrication capabilities, software and computing.

**Recommendation 5.2:** The ESA mandatory science programme covers projects supporting almost the entirety of the UK astronomy science and technology communities, with missions comparable (or greater) in cost, impact and community support to our highest-rated ground-based projects. The UK must continue ESA membership, and there is widespread astronomy community appetite for exploring bilateral space agreements and other space opportunities.

**Recommendation 5.3:** Maintaining UK access to the top priority facilities is not sufficient to ensure returns on these major investments, so STFC should maintain and develop a broad portfolio of high/very high priority science facilities, including (but not limited to) WEAVE, LOFAR, JCMT, GOTO, LT/NRT, WASP, NGTS, DIRAC, IRIS, e-MERLIN, Lovell Telescope, among current facilities, and UK roles in e.g. DESI, CCAT, 4MOST, BlackGEM, as well as fostering roles in currently non-STFC-supported projects such as REACH; looking further ahead, HARPS-3, Rubin LSST, Simons Observatory, AtLAST, MSE, CTA, VLT/MOONS. As part of this broad portfolio supporting the highest priority science, AAP further recommends support for laboratory astrochemistry.

# 8. Appendix: List of acronyms

4MOST: 4-metre Multi-Object Spectroscopic Telescope

AAP: Astronomy Advisory Panel

AGP: Astronomy Grants Panel

ALMA: Atacama Large Millimetre Array

ANDES: ArmazoNes high Dispersion Echelle Spectrograph (formally known as HIRES)

AO: Adaptive Optics

APEX: Atacama Pathfinder Experiment

APPEC: Astroparticle Physics European Consortium

ASTRONET: a network of European funding agencies and research organisations in astronomy (not an acronym, but capitalised)

ASTRI: Astrofisica con Specchi a Tecnologia replicante Italiana

ATC: Astronomy Technology Centre

AtLAST: Atacama Large Aperture Submillimeter Telescope

BINGO: Baryon Acoustic Oscillations from Integrated Neutral Gas Observations

BiSON: Birmingham Solar Oscillations Network

BISOU: Balloon Interferometer for Spectral Observations of the Universe

BlackGEM: a wide-field array of optical telescopes to be located at ESO's La Silla Observatory in Chile's Atacama desert (not an acronym[58])

BlueMUSE: Blue Multi Unit Spectroscopic Explorer

BoP: Balance of Programmes

CAHA: Centro Astronómico Hispano-Alemán

CARMENES: Calar Alto high-Resolution search for M dwarfs with Exoearths with Near-infrared and optical Échelle Spectrographs

CBASS: C-Band All Sky Survey

CCAT: Cerro Chajnantor Atacama Telescope

CCD: Charge Coupled Device

CFHT: Canada-France-Hawaii Telescope

CHARA: Center for High Angular Resolution Astronomy

CMB: Cosmic Microwave Background

---

[58] See https://www.eso.org/public/about-eso/acronyms/

CMB-S4: Stage Four Cosmic Microwave Background experiment

CMOS: Complementary metal–oxide–semiconductor

Co-I: Co-investigator

CONCERTO: Carbon CII line in post-reionisation and reionisation epoch

CRIRES, CRIRES+: Cryogenic high-resolution cross-dispersed infrared echelle spectrograph

CSA: Canadian Space Agency

CTA: Cherenkov Telescope Array

CUBES: Cassegrain U-Band Efficient Spectrograph

CUGAS: Consortium of Universities for Goonhilly Astronomy and Space

DESI: Dark Energy Spectroscopic Instrument

DiRAC: Distributed Research using Advanced Computing

DKIST: Daniel K. Inouye Solar Telescope

EAGLE: Evolution and Assembly of GaLaxies and their Environments

EC: European Commission

EDI: Equity, Diversity and Inclusion

EHT: Event Horizon Telescope

ELT: Extremely Large Telescope

e-MERLIN: enhanced Multi Element Remotely Linked Interferometer Network

EOSC: European Open Science Cloud

ERIC: European Research Infrastructure Consortium

ERIS: Enhanced Resolution Imager and Spectrograph

ESA: European Space Agency

ESO: European Southern Observatory

ESPRESSO: Echelle SPectrograph for Rocky Exoplanets and Stable Spectroscopic Observations

EVN: European VLBI Network

FAIR: Findable, Accessible, Interoperable, Reusable

FRB: Fast Radio Burst

GMRT: Giant Metrewave Radio Telescope

GNOSIS: Global Network on Sustainability in Space

GOTO: Gravitational-wave Optical Transient Observer

GRAND: Giant Radio Array for Neutrino Detection

GranTeCan: Gran Telescopio Canarias

GRAVITY, GRAVITY+: a 4-beam interferometric combiner at VLTI operating in K-band (not an acronym[59])

GRST: Fermi Gamma-Ray Space Telescope

GTM: Gran Telescopio Milimétrico Alfonso Serrano

GW: Gravitational Wave

HARMONI: High Angular Resolution Monolithic Optical and Near-infrared Integral field spectrograph

HARPS, HARPS-3: High Accuracy Radial velocity Planet Searcher

HARPS-N: High Accuracy Radial velocity Planet Searcher, North

HERA: Hydrogen Epoch of Reionization Array

HESS: High Energy Stereoscopic System

HIRES: high-resolution spectrograph (ELT instrument now known as ANDES)

HiRISE: High-Resolution Imaging and Spectroscopy of Exoplanets

HPC: High Performance Computing

IMF: Initial Mass Function

ING: Isaac Newton Group

INT: Isaac Newton Telescope

IRIS: eInfrastructure for Research and Innovation in STFC

JCMT: James Clerk Maxwell Telescope

JIVE: Joint Institute for VLBI ERIC

JWST: NASA/ESA/CSA next-generation infrared space telescope. (Following the provisional best practice outlined by the Royal Astronomical Society[60], we do not expand the acronym.)

JVLA: Jansky Very Large Array

KAGRA: Kamioka Gravitational Wave Detector

KIDS: Kinetic Inductance Detectors

LGBTQI+: Lesbian, Gay, Bisexual, Transgender, Queer, Intersex

LHC: Large Hadron Collider

LIGO: Laser Interferometer Gravitational-Wave Observatory

LISA: Laser Interferometer Space Antenna

LMT: Large Millimetre Telescope (in Spanish: GTM, Gran Telescopio Milimétrico Alfonso Serrano)

LOFAR: Low Frequency Array

LSPE-strip: Large-Scale Polarization Explorer/Survey Tenerife Polarimeter

---

[59] See https://www.eso.org/public/about-eso/acronyms/
[60] See https://ras.ac.uk/news-and-press/news/ras-and-jwst

LSPE-swipe: Large-Scale Polarization Explorer/Short Wavelength Instrument for the Polarization Explorer

LSST: Legacy Survey of Space and Time

LT: Liverpool Telescope

MAGIC: Major Atmospheric Gamma Imaging Cherenkov telescopes

MATISSE: Multi AperTure mid-Infrared SpectroScopic Experiment

MAVIS: MCAO-Assisted Visible Imager and Spectrograph

MCAO: Multi-Conjugate Adaptive Optics

MeerKAT: originally the Karoo Array Telescope (not an acronym)

MeerLICHT: the Dutch translation for 'more light', is an astronomical project which aims to provide a simultaneous, real-time optical view of the radio (transient) sky as observed by MeerKAT. (Not an acronym)

MeerTime: a five-year pulsar timing program on the MeerKAT array (not an acronym)

METIS: Mid-Infrared ELT Imager and Spectrograph

MICADO: Multi-AO Imaging Camera for Deep Observations

MIDI: Mid-infrared interferometric instrument

MKIDS: Microwave Kinetic Inductance Detectors

MOSAIC: Multi-Object Spectrograph (for ELT)

MOONS: Multi-Object Optical and Near-infrared Spectrograph

MROI: Magdalena Ridge Observatory Interferometer

MSE: Mauna Kea Spectroscopic Explorer

MUSE: Multi Unit Spectroscopic Explorer

MWA: Murchison Widefield Array

NASA: National Aeronautics and Space Administration

NEID: NN-explore Exoplanet Investigations with Doppler spectroscopy

ngEHT: Next generation Event Horizon Telescope

NGTS: Next-Generation Transit Survey

ngVLA: Next Generation Very Large Array

NIR: Near Infra-Red

NIRPS: Near Infra Red Planet Searcher

NN-Explore: NASA-NSF Explore, the joint exoplanet programme of NASA and the NSF

NRT: New Robotic Telescope

NSF: National Science Foundation

NTT: New Technology Telescope

Page 35 of 37

PAAP: Particle Astrophysics Advisory Panel

PCS: Planetary Camera and Spectrograph

PDRA: Post-Doctoral Research Associate

PI: Principal Investigator

PFS: Prime Focus Spectrograph

PPRP: Projects Peer Review Panel

QUBIC: Q-U Bolometric Interferometer for Cosmology

QUIJOTE: Q-U-I JOint TEnerife CMB experiment

REACH: Radio Experiment for the Analysis of Cosmic Hydrogen

RISTRETTO: high-Resolution Integral-field Spectrograph for the Tomography of Resolved Exoplanets Through Timely Observations[61]

SDO: Solar Dynamics Observatory

SDSS: Sloan Digital Sky Survey

SDSS-V: the fifth Sloan Digital Sky Survey

SKA: Square Kilometer Array

SO: Simons Observatory

SPHERE: Spectro-Polarimetic High contrast imager for Exoplanets REsearch

SPIROU: SPectropolarimètre InfraROUge

SSAP: Solar System Advisory Panel

STFC: Science and Technology Facilities Council

TRL: Technology Readiness Level

TNG: Telescopio Nazionale Galileo

uGMRT: upgraded Giant Metrewave Radio Telescope

UK: United Kingdom

UKNRAO: UK National Radio Astronomical Observatory

UKRI: UK Research and Innovation

UKSA: UK Space Agency

US: United States

UV: Ultraviolet

UVES: Ultraviolet and Visual Echelle Spectrograph

WASP: Wide-Angle Search for Planets

WEAVE: WHT Enhanced Area Velocity Explorer

---

[61] See e.g. https://zenodo.org/record/3356296

WHT: William Herschel Telescope

WIYN: Telescope operated by the University of Wisconsin–Madison (W), Indiana University (I), Yale University (Y), and the National Optical Astronomy Observatories (N).

VISTA: Visible and Infrared Survey Telescope for Astronomy

VLA: Very Large Array

VLBI: Very Long Baseline Interferometry

VLT: Very Large Telescope

VLTI: Very Large Telescope Interferometer